\documentclass[numsec,webpdf,modern,medium,namedate]{oup-authoring-template}

\onecolumn 

\graphicspath{{Fig/}}

\theoremstyle{thmstyleone}%
\theoremstyle{thmstyletwo}%
\theoremstyle{thmstylethree}%

\usepackage{graphicx} 
\usepackage{bm}
\usepackage{amsmath}
\usepackage{amssymb}
\usepackage{color}
\usepackage{algorithm}
\usepackage{algpseudocode}

\begin{document}

\journaltitle{\textit{arXiv}}
\DOI{DOI HERE}
\copyrightyear{2025}
\pubyear{---}

\firstpage{1}


\title[Profile-LMM]{Bayesian Profile Regression with Linear Mixed Models (Profile-LMM) applied to Longitudinal Exposome Data}

\author[1,2\,$\ast$]{Matteo Amestoy\ORCID{0000-0002-2032-1448}}
\author[1,2]{Mark van de Wiel}
\author[1,2,3]{Jeroen Lakerveld}
\author[1,2,4]{Wessel van Wieringen}

\authormark{Amestoy et al.}

\address[1]{\orgdiv{Department of Epidemiology \& Data Science}, \orgname{Amsterdam UMC}, \orgaddress{\state{Amsterdam}, \country{The Netherlands}}}
\address[2]{\orgdiv{Amsterdam Public Health research institute},\orgname{Amsterdam UMC}, \orgaddress{\state{Amsterdam}, \country{The Netherlands}}}
\address[3]{\orgdiv{Upstream Team}, \orgname{Amsterdam UMC}, \orgaddress{\state{Amsterdam}, \country{The Netherlands}}}
\address[4]{\orgdiv{Department of Mathematics}, \orgname{Vrije Universiteit Amsterdam}, \orgaddress{\state{Amsterdam}, \country{The Netherlands}}}

\corresp[$\ast$]{Corresponding author. \href{email:email-id.com}{m.amestoy@amsterdamumc.nl}}




\abstract{Exposure to diverse non-genetic factors, known as the exposome, is a critical determinant of health outcomes. However, analyzing the exposome presents significant methodological challenges, including: high collinearity among exposures, the longitudinal nature of repeated measurements, and potential complex interactions with individual characteristics. In this paper, we address these challenges by proposing a novel statistical framework that extends Bayesian profile regression. Our method integrates profile regression, which handles collinearity by clustering exposures into latent profiles, into a linear mixed model (LMM), a framework for longitudinal data analysis. This profile-LMM approach effectively accounts for within-person variability over time while also incorporating interactions between the latent exposure clusters and individual characteristics. We validate our method using simulated data, demonstrating its ability to accurately identify model parameters and recover the true latent exposure cluster structure. Finally, we apply this approach to a large longitudinal data set from the Lifelines cohort to identify combinations of exposures that are significantly associated with diastolic blood pressure.}


\maketitle

\section{Introduction}

The concept of the exposome \citep{wild_complementing_2005}, which encompasses the totality of non-genetic factors, has emerged as a critical framework for understanding complex disease etiologies. Within this framework, a central challenge in contemporary biomedical science is to dissect the intricate relationship between diverse environmental and lifestyle factors (exposures) and their influence on health outcomes. In this paper, we focus on modeling the relationship between diastolic blood pressure (DBP) and multiple exposures. This focus is motivated by the fact that hypertension is identified as one of the leading global risk factors for mortality \citep{murray_global_2020}, with extensive evidence linking blood pressure regulation to environmental factors \citep{arku_long-term_2020,nieuwenhuijsen_influence_2018,munzel_heart_2021}. Our analysis utilizes the rich longitudinal data from the Lifelines cohort \citep{sijtsma_cohort_2022}, which integrates individual-level characteristics from three waves of measurements with a detailed suite of exposure metrics, including pollutant densities at residential addresses and neighborhood livability measures.\\
Our inquiry is complicated by several methodological challenges. First, the inherent collinearity among exposure metrics, such as home-address pollutants, makes it difficult to disentangle the independent contribution of each factor using conventional regression models. Second, the repeated measurements setting necessitates an approach that can appropriately model the longitudinal dynamics and within-person variability over time. Finally, we want to take into account the potential interactions between exposures and individual-level characteristics, which are incorporated into the analysis to achieve a comprehensive understanding. These challenges motivate the development of novel approaches to effectively analyze exposome data.\\
To address the challenge of collinearity, Bayesian profile regression, as introduced by \cite{molitor_bayesian_2010}, offers a powerful framework. This method jointly performs two steps: it first clusters a subset of covariates (in our case, the exposures) into a set of interpretable clusters and then uses the resulting cluster memberships in a regression model. This approach is particularly effective at identifying distinct, easily interpretable groups of exposures and their combined influence on health outcomes. While profile regression has been successfully applied in cross-sectional exposome studies \citep{coker_multi-pollutant_2018,griffin_assessment_2022}, its applicability is limited to static data. A recent extension by \cite{rouanet_bayesian_2023} adapted profile regression for longitudinal data, but their work is limited to identifying latent clusters of individuals, rather than exposures, sharing temporal trajectories. Moreover, it does not account for potential interactions between exposure clusters and individual characteristics.\\
To overcome these limitations, we introduce a novel extension of Bayesian profile regression by integrating it into a Linear Mixed Model (LMM) framework. LMMs are a versatile and computationally efficient tool for analyzing longitudinal data, particularly to account for variability within the person. Our proposed profile-LMM approach effectively combines the strengths of both methods: it preserves the ability of profile regression to identify latent clusters of exposures that affect the outcome, while leveraging the LMM's capacity to handle repeated measures. Crucially, this framework also allows for the seamless incorporation of interactions between the latent exposure clusters and individual-level characteristics, providing a more comprehensive and biologically plausible model of the exposome-health relationship.\\
The remainder of the paper is organized as follows. In Section \ref{sec:model}, we recall the original formulation of Bayesian profile regression and present our LMM extension and the resulting inference strategy. Section \ref{sec:Simulation} validates our approach using a simulated dataset, demonstrating the method's ability to correctly estimate the parameters of the LMM and identify the latent cluster structure. Finally, in Section \ref{sec:RealData}, we apply our method to the Lifelines cohort dataset to identify meaningful combinations of exposures that influence DBP.

\section{Bayesian linear mixed model profile regression}\label{sec:model}

\subsection{Bayesian profile regression}

Profile regression was introduced by \cite{molitor_bayesian_2010}, and models outcomes influenced by complex covariate structures. Fundamentally, profile regression explains an outcome variable, $\mathbf{y}$, based on a set of explanatory covariates that decomposes into two distinct groups. The first set of covariates, denoted as $\mathbf{u}$, comprises strongly correlated variables (e.g., questionnaire responses in the original paper of \cite{molitor_bayesian_2010}, exposure variables in our case) for which direct incorporation into the regression is likely far from optimal due to interdependence. The second set of covariates, $\mathbf{x}$, consists of conventional explanatory variables (e.g. confounding individual characteristics in both the authors and our study). To address the challenges posed by $\mathbf{u}$, \cite{molitor_bayesian_2010} propose to cluster these covariates and employ the resulting cluster memberships as indicator variables, in conjunction with $\mathbf{x}$, in the regression model. Unlike traditional methods that perform clustering and regression sequentially, profile regression simultaneously estimates the parameters for both models. Consequently, we will refer to $\mathbf{x}$ as regression covariates and $\mathbf{u}$ as clustering covariates in the remainder of the paper.\\

Formally, profile regression considers a dataset comprising $n$ observations of the form $\mathcal{D} = \{(y_i, \mathbf{u}_i,\mathbf{x}_i) \mid i = 1, ..., n\}$. Each observation $i$ consists of an outcome variable, $y_i$ (e.g., health outcome), a corresponding vector of clustering covariates, $\mathbf{u}_i$ (e.g., exposure variables) and a vector of regression covariates $\mathbf{x}_i$, (e.g., individual characteristics). Unlike conventional regression models that focus on the conditional distribution of the outcome given all the covariates, Bayesian profile regression models the joint distribution of $(y_i, \mathbf{u}_i)$ conditionally on $\mathbf{x}_i$. A mixture model is employed to represent this joint distribution. The likelihood function of this model is expressed as:
\begin{equation}\label{eq:main}
p(y_i,\mathbf{u}_i|\bm{\pi},\bm{\theta},\mathbf{x}_i,\bm{\alpha}) = \sum_{c=1}^{C}\pi_c f(y_i, \mathbf{u}_i | \bm{\theta}_c,\mathbf{x}_i,\bm{\alpha}) = \sum_{c=1}^{C}\pi_c f_y(y_i | \bm{\theta}_c^y,\mathbf{x}_i,\bm{\alpha}) f_u(\mathbf{u}_i | \bm{\theta}_c^u),  
\end{equation}

where $f(.|\bm{\theta}_c,\mathbf{x}_i,\bm{\alpha})$ is the mixture component density, decomposable into $f_y(y_i | \bm{\theta}_c^y,\mathbf{x}_i,\bm{\alpha})$ representing the outcome or regression distribution and $f_u(\mathbf{u}_i | \bm{\theta}_c^u)$ representing the assignment distribution. The second equality assumes that the outcome $y_i$ and covariates $\mathbf{u}_i$ are independent, conditionally on the mixture component specific parameters $\bm{\theta}_c=\{ \bm{\theta}_c^u, \bm{\theta}_c^y\}$. The vector $\bm{\alpha}$ is the population constant parameters shared by all the observations and $\bm{\pi} = \{\pi_c\}$ represents the mixture weights. Consistent with the model's original formulation, the number of mixtures, $C$, is treated as infinite. The computational challenge of this infinite state space is addressed in detail in Section \ref{sec:Inference}.\\

The selection of outcome and assignment functions depends on the specific problem domain, considering the nature of the covariates and outcomes. For instance, \cite{molitor_bayesian_2010} focused on a scenario where clustering covariates were comprised of $q$ binary variables. They employed independent Bernoulli distributions for the assignment function, $f_u(\mathbf{u}_i | \bm{\theta}_c^u) = \prod_{j=1}^q ( \theta_{c,j}^u)^{u_{i,j} } ( 1-\theta_{c,j}^u)^{(1-u_{i,j} )}$. Here, $\bm{\theta}_c^u = \{\theta_{c,j}^u\}_{j = 1\dots q}$ and $\theta_{c,j}^u$ represents the probability that the $j^\text{th}$ clustering covariate is present in mixture $c$.  In terms of the outcome of interest, a binary variable, they adopted a logistic regression model, $f_y(y_i | \bm{\theta}_c^y,\mathbf{x}_i,\bm{\alpha}) = \text{logit}\big(p(y_i = 1)\big) = \mathbf{x}_i\bm{\alpha} +\theta_c^y$, incorporating a mixture-specific constant  $\theta_c^y$. \\

\subsection{Linear mixed model profile regression}

In this subsection, we will detail the types of outcome and assignment distributions that our proposed model can accommodate.\\

We begin with the regression distribution, $f_y(y_i | \bm{\theta}_c^y,\mathbf{x}_i,\bm{\alpha}) $, which constitutes the innovative aspect of our proposed method. Indeed,  Bayesian profile regression has primarily been applied to outcomes derived from cross-sectional data (see \cite{molitor_bayesian_2010,papathomas_exploring_2012} and the corresponding R package \cite{liverani_premium_2015}). Within this context the standard method and its extensions offer versatility in accommodating diverse outcome types, including categorical, count, and continuous data. Recent advancements by \cite{rouanet_bayesian_2023} extended the original model to encompass longitudinal continuous data. However, their approach exhibits limitations that render it unsuitable for our specific problem. As we elucidate our model we show that their approach can be considered a special case of our proposed framework.\\

Our approach uses LMMs as the outcome model. LMMs represent a more general framework for analyzing continuous longitudinal data, offering superior flexibility and the ability to account for the hierarchical structure of the data.\\
We first consider a simplified scenario for the regression distribution, focusing on the longitudinal data part of the model, without the influence from the cluster mixture components $\bm{\theta}_c^y$. Assuming, without loss of generality, that the data comprise repeated measurements collected from a set of $m$ individuals, we introduce a mapping function $g: [1,n] \rightarrow [1,m]$ that associates each observation index $i \in [1,n]$ with the corresponding individual index, thereby linking each observation to its originating individual. For each observation $i$, we observe a continuous outcome variable $y_i$ along with a set of covariates $\mathbf{x}_i$. A common approach to model such data is an LMM and is expressed as follows:

\begin{align*}
    y_i &= \mathbf{x}^{\text{Fe}}_i\bm{\beta} + \mathbf{x}^{\text{Re}}_i\bm{\eta}_{g(i)}+\epsilon_i,\\
    \bm{\eta}_{j}&\sim\mathcal{N}(0,\mathbf{W}^{\text{Re}}),\,\,\, \text{for all }j \in [1,m] \\
    \epsilon_i &\sim\mathcal{N}(0,\sigma^2)
\end{align*}
where the vectors $\mathbf{x}^{\text{Fe}}_i$ and $\mathbf{x}^{\text{Re}}_i$ represent subsets of the covariates $\mathbf{x}_i$, encompassing fixed effect covariates and random effect covariates, respectively. In the particular case where $\mathbf{x}^{\text{Re}}_i$ is a constant, we are dealing with a random intercept model, whereas if $\mathbf{x}^{\text{Fe}}_i=\mathbf{x}^{\text{Re}}_i$, all covariates are associated with random slopes.
The vector $\bm{\beta}$ denotes the fixed effect coefficients, which remain constant across all observations, while $\bm{\eta}_{g(i)}$ represents the random effect parameters shared by all observations originating from individual $g(i)$. The random effects are assumed to follow a Gaussian distribution with a constant random effect covariance matrix $\mathbf{W}^{\text{Re}}$. Lastly, the error term $\epsilon_i$ is homoscedastic with variance $\sigma^2$.\\

We now incorporate the mixture-specific  component, $\bm{\gamma}_c$, into the conditional regression model of equation (\ref{eq:main}), $f_y(y_i | \bm{\theta}_c^y,\mathbf{x}_i,\bm{\alpha}) $:
\begin{align}\label{eq::LMM}
    y_i &= \mathbf{x}^{\text{Fe}}_i\bm{\beta} + \mathbf{x}^{\text{Re}}_i\bm{\eta}_{g(i)}+\mathbf{x}^{\text{Int}}_i\bm{\gamma}_c+\epsilon_i,\\
    \bm{\eta}_{j}&\sim\mathcal{N}(0,\mathbf{W}^{\text{Re}}),\,\,\, \text{for all }j \in [1,m]\nonumber \\
    \epsilon_i &\sim\mathcal{N}(0,\sigma^2)\nonumber
\end{align}
where the vector $\bm{\gamma}_c$ is a cluster mixture-specific parameter. The vector $\mathbf{x}^{\text{Int}}_i$ represents a subset of the covariates $\mathbf{x}_i$, that are hypothesized to interact with the clustering covariates $\mathbf{u}_i$ (exposure in our case). If $\mathbf{x}^{\text{Int}}_i$ is a constant intercept, then the effect of the exposures on the outcome is limited to a cluster mixture specific offset. To summarize, $\mathbf{x}^{\text{FE}}_i,\mathbf{x}^{\text{RE}}_i,\mathbf{x}^{\text{Int}}_i$ are all (potentially overlapping) subsets of the regression covariates $\mathbf{x}_i$, that denote the part of the covariates that respectively compose the fixed effects, the individual random effects and the variables that interact with the latent exposure clusters.
Employing the notations from the previous section, we define the population parameters as $  \bm{\alpha} = \{\bm{\beta},\sigma^2,\mathbf{W}^{\text{Re}}\}$ and the mixture parameters as $ \bm{\theta}_c^y = \bm{\gamma}_c$.\\

 We now turn our attention to briefly outlining the assignment distribution of equation (\ref{eq:main}), $f_u(\mathbf{u}_i | \bm{\theta}_c^u)$, adhering to the structure proposed by \cite{liverani_premium_2015}. The model is capable of accommodating both continuous and categorical clustering covariates $\mathbf{u}_i$.  In the case of categorical covariates, $ \mathbf{u}_i= (u_{i,1},\cdots, u_{i,q^u})$, each component is assumed to follow an independent multinomial distribution, $u_{i,j}|\bm{\theta}_c^u\sim\mathcal{M}(\bm{\phi}_{c,j})$. The vector $\bm{\phi}_{c,j}$ represents the probabilities associated with the categories of $u_{i,j}$. For continuous covariates, a multivariate Gaussian distribution is employed, $\mathbf{u}_i|\bm{\theta}_c^u \sim \mathcal{N}(\bm{\mu}_c,\bm{\Sigma}_c)$. In the case where the covariate vector $ \mathbf{u}_i$  comprises a mixture of continuous variables and $l$ categorical variables, we assume independence between these two types and partition the assignment parameters accordingly and $\bm{\theta}_c^u = \{\bm{\phi}_{c,1},\cdots,\bm{\phi}_{c,l},\bm{\mu}_c,\bm{\Sigma}_c\}$.\\

Our proposed model shares similarities with the work of \cite{rouanet_bayesian_2023} in its adaptation of profile regression to longitudinal data. However, a key distinction lies in the clustering objective: while \cite{rouanet_bayesian_2023} focus on clustering individuals, our model clusters observations based on their exposure levels. In our application, observations from the same individual may belong to different latent clusters if their exposure levels varied over time, such as in the case of a residential move between measurement periods. This approach effectively decouples repeated measurement correlation modeling from latent profile cluster estimation. 

Furthermore, \cite{rouanet_bayesian_2023} restricts the influence of mixture components on the outcome to time functions. While their method is not directly applicable to our specific problem, our model can be adapted to address their scenario, demonstrating its broader applicability. In fact, our proposed model can be viewed as a generalization of \cite{rouanet_bayesian_2023}'s model. 

Currently, our model is formulated at the observation level, motivated by our practical application. However, with minor modifications, it can be adapted to the individual level. By redefining the index $i$ to represent individuals rather than observations, $\mathbf{y}_i$ becomes a vector of repeated measurements for individual $i$, and $\mathbf{x}_i$ becomes a matrix of covariates for each measurement.

With this adjustment, our model aligns more closely with \cite{rouanet_bayesian_2023}'s formulation. Their model expresses the outcome for individual $i$ as:

\begin{align*}
\mathbf{y}_i &= \mathbf{x}^{\text{Fe}}_i\bm{\beta} + f_c(t_i)+\epsilon_i,\\
 \epsilon_i &\sim\mathcal{N}(0,\sigma^2),
\end{align*}
where $f_c$ is a mixture-specific Gaussian process. While our method may not achieve the full flexibility of a general Gaussian process, it can approximate similar behavior by incorporating temporal basis functions (e.g., splines) within the covariate matrix $\mathbf{x}^{\text{Int}}$. However, our approach offers additional advantages, including the ability to explore interactions between latent classes and other covariates beyond time. Moreover, the model proposed by \cite{rouanet_bayesian_2023} relies solely on latent classes to capture shared information across observations. Their approach inherently lacks control over potential within-individual dependencies across repeated measurements. In our proposed outcome model, the random effect variable captures the shared information across observations at the individual level.\\

\subsection{Inference}\label{sec:Inference}
Within a Bayesian framework, we employ Markov chain Monte Carlo (MCMC) methods to infer the posterior distribution of model parameters. To enhance computational efficiency, given the intensive nature of MCMC and the potential high dimensionality of the data, our choices of priors are guided by practical considerations.\\
This subsection is structured as follows: we commence by specifying prior distributions for all model parameters. Subsequently, we outline the general structure of the Gibbs sampling algorithm utilized to obtain posterior samples. Finally, we elucidate the process of parameter estimation from the posterior samples.

\subsubsection{Parameters and priors}
In the subsequent section, we will specify our chosen prior distributions for the mixture components, followed by those for the assignment distribution and, finally, those for the outcome distribution.\\

As is customary with infinite mixture models, a Dirichlet process prior is specified. Following \cite{liverani_premium_2015}, a stick breaking process prior \citep{ishwaran_gibbs_2001,pitman_combinatorial_2006} is imposed on the mixture weights of equation (\ref{eq:main}), $\pi_c$:
\begin{align*}
    V_c&\sim \beta(1,\zeta),\, \text{ i.i.d for } c\in\mathbb{Z}^+\\
    \pi_1 &= V_1,\,\,\,\,\pi_c = V_c\prod_{j=1}^{c-1}(1-V_j)
\end{align*}
Here, $\beta(1,\zeta)$ denotes the beta distribution with shape parameters 1 and $\zeta$. The parameter $\zeta$  governs the expected number of mixture components and is consequently estimated within the model using a gamma prior. \\
To enhance computational tractability, we adopt a truncated Dirichlet process mixture model (DPMM) framework, as originally proposed by \cite{molitor_bayesian_2010}. This approach assumes a finite maximum number of clusters, effectively restricting the latent parameter space to a finite dimension.  When the maximum number of clusters is sufficiently large, the truncation leads to a model that is virtually indistinguishable from one based on an infinite-dimensional prior \citep{ishwaran_gibbs_2001}. Compared to standard profile regression, this approach eliminates the need for a Metropolis within Gibbs step, required to ensure adequate mixing properties of the sampled cluster allocations.\\

We now specify the prior distributions for the cluster assignment distribution. For each categorical covariate $u_{i,j}$ comprising $\mathbf{u}_{i}$, a conjugate Dirichlet prior is adopted for $\bm{\phi}_{c,j}$. For the set of continuous covariates composing $\mathbf{u}_{i}$, a conjugate normal-inverse-Wishart prior is specified for $(\bm{\mu}_c,\bm{\Sigma}_c)$. In the case where the covariate vector $\mathbf{u}_i$ comprises a mixture of continuous variables and categorical variables, their respective parameters are endowed with associated independent priors.\\

For the outcome distribution, we adhere to the standard conjugate setup. A normal-gamma prior is adopted on the fixed effect parameter and error variance term $(\bm{\beta},\sigma^2)$, and an inverse-Wishart prior is specified for the variance matrix of the random effects $\mathbf{W}^{\text{Re}}$.
Regarding the latent clusters regression parameters $ \gamma_c$, we employ a Gaussian prior with covariance $\mathbf{W^{\text{Int}}}$.
The influence of the latent factor matrix, $\mathbf{W^{\text{Int}}}$, can be substantial and is generally unknown \emph{a priori}. To address this challenge, we estimate $\mathbf{W^{\text{Int}}}$ by incorporating it into the model, with an inverse-Wishart prior.

\subsubsection{Posterior sampling with a Gibbs sampler}

We employ a Markov chain Monte Carlo (MCMC) method to sample from the posterior distribution of the model parameters. These parameters include:
\begin{itemize}
    \item outcome distribution parameters: $\{\bm{\alpha},\bm{\theta}_1^y,\dots,\bm{\theta}^y_C,\boldsymbol{W}^\text{Int}\}$, comprising population constant parameters $ \bm{\alpha} = \{\bm{\beta},\sigma^2,\boldsymbol{W}^\text{Re}\}$, mixture-dependent parameters $\bm{\theta}^y_c = \bm{\gamma}_c$ and $\boldsymbol{W}^\text{Int}$ the parameters of the mixture dependent parameters' prior. 
\item asssignment distribution mixture dependent parameters: $\bm{\theta}^u = \{\bm{\theta}_1^u,\dots,\bm{\theta}^u_C\}$, where $\bm{\theta}_c^u = \{\bm{\phi}_{c,1},\cdots,\bm{\phi}_{c,l},\bm{\mu}_c,\bm{\Sigma}_c\}$ ,
\item mixture parameters: $\bm{\theta}^m = \{\bm{\pi},\zeta\}$.
\end{itemize}
We employ a Gibbs sampler to sample from the posterior distribution. As is common practice  \citep{liverani_premium_2015} in the context of mixture model inference, we introduce a vector of latent allocation variables, denoted by $\mathbf{Z}$. For each observation $i$, the latent variable $\mathbf{Z}_i$ dictates the mixture component associated to observation $i$, such that $$p(y_i,\mathbf{u}_i|\bm{\theta},\mathbf{Z}_i=c) = f(y_i,\mathbf{u}_i|\bm{\theta}_c)\text{ and } p(\mathbf{Z}_i=c|\bm{\pi})=\pi_c.$$ In essence, $\mathbf{Z}_i$ represents a cluster membership, where all observations sharing the same $\mathbf{Z}_i$ value are assigned to the same mixture component and consequently share the same underlying distribution parameters.\\

By incorporating $\mathbf{Z}$ into the parameter space, the posterior distributions of the outcome and assignment parameters become conditionally independent given $\mathbf{Z}$. As each block represents a relatively simple model, we can derive an efficient Gibbs sampler.

A detailed step-by-step description of the algorithm is presented in the Appendix. The design of our Gibbs sampler is largely inspired by \cite{ishwaran_gibbs_2001}, which provides a comprehensive discussion of Dirichlet process mixture model sampling, and \cite{liverani_premium_2015}, which delves into the details of profile regression model inference. 

Convergence, or lack thereof, in such mixture models can be assessed by examining the trace plots of global parameters, such as $\zeta$ and the number of non-empty clusters.\\

\subsubsection{Inference and post-processing of the posterior samples}

While posterior inference for population-wide parameters is readily achieved by computing summary statistics from the empirical posterior distribution, extracting meaningful information from the latent parameters presents a greater challenge. This is due to the well-known label switching problem, where the cluster labels are not identifiable and can permute across MCMC samples. Each MCMC sample contains a valid set of proposed cluster assignments and corresponding cluster characteristics, but the lack of consistent labeling across iterations prohibits the direct aggregation of cluster-specific parameters. \\
To address this issue, we follow the solution proposed in \cite{molitor_bayesian_2010,liverani_premium_2015}. This approach involves constructing a representative clustering that summarizes the latent information captured throughout the MCMC chain.

The key step involves summarizing the entire sampled allocation chain into an average dissimilarity matrix, denoted by $\mathbf{S}$. This matrix represents the proportion, across the MCMC chain, of samples where a pair of observations were assigned to the same cluster. Subsequently, a Partitioning Around Medoids (PAM) algorithm is employed using the average dissimilarity matrix to identify a representative clustering (denoted by $\mathbf{Z}^{\ast}$).

Once the representative clustering is obtained, we can estimate the cluster-specific parameters and their associated uncertainties by integrating over the entire MCMC chain, effectively leveraging the information from all the sampled latent configurations. 

We demonstrate the methodology using the outcome mixture specific parameter $\bm{\gamma}$, recognizing that the same procedure is applicable to the cluster-specific assignment distribution mixture $\bm{\theta}^u$. For each cluster $c$ identified by the representative clustering $\mathbf{Z}^{\ast}$, we compute the mean a posteriori estimate, $\bm{\gamma}^\ast_c$, of the parameter by aggregating the posterior samples across both MCMC iterations and all the observations assigned to cluster $c$. The formula for this mean estimate is: $$ \bm{\gamma}^\ast_c = \frac{1}{HN_c}\sum_{h=1}^H\sum_{i|\mathbf{Z}^{\ast}_i=c}\bm{\gamma}^{(h)}_{\mathbf{Z}^{(h)}_i} $$ Here, $\bm{\gamma}^{(h)}_{\mathbf{Z}^{(h)}_i}$ denotes the sampled parameter for the $h^{\text{th}}$ MCMC sample associated with observation $i$. $H$ represents the total number of samples retained in the chain, and $N_c$ is the number of observations such that $\mathbf{Z}^{\ast}_i=c$. In effect, the posterior sample pool for the parameter of a given cluster $c$ is formed by aggregating all parameters $\bm{\gamma}^{(h)}_{i}$ across all $H$ MCMC samples and all $N_c$ observations for which the corresponding observation $i$ is assigned to that cluster. Moreover, credible intervals and other standard Bayesian statistical measures can be computed using analogous methodologies.\\

\section{Simulation study}\label{sec:Simulation}

In this section, we demonstrate the efficacy of our method in two variations of an experimental setting inspired by real-world exposure data. Our primary objectives are to evaluate the precision of the LMM's parameter inference and the capacity of our method to identify the exposure latent cluster structure.

\subsection{Design and scenarios}
Consider a longitudinal study of the evaluation of exposure on health. Following the design of our real-world data application, we examine a third-wave longitudinal study involving 1000 individuals, resulting in 3000 total observations. To simulate the random nature of observation times, for each individual, the timing of the $i^{th}$ wave is drawn uniformly from the interval $[i, i+1)$. At each observation time, a continuous outcome variable $\mathbf{Y}$ and four individual characteristics $\mathbf{x}$ are measured. These individual characteristics exhibit varying properties, including time-invariant, binary, and continuous variables, mirroring the diversity encountered in real-world studies. In addition, two exposure variables, $\mathbf{u} = [\mathbf{u}_1, \mathbf{u}_2]$, are measured at each time point.\\

We assume that the two exposure variables define nine clusters corresponding to a Gaussian mixture model (GMM) with centroids located on a mesh grid defined by the set $\{-1, 0, 1\}$. A positive exposure value is interpreted as indicative of a healthier environment. Each of the nine unique combination of negative, positive, or neutral exposure values is hypothesized to have a distinct impact on the health outcome.  

Given these clusters, we propose a linear mixed-effects profile regression model to capture the relationship between the outcome variable, $\mathbf{Y}$, and the covariates. Following equation (\ref{eq::LMM}), our model incorporates the following components:
\begin{itemize}
    \item Fixed Effects: The fixed-effects component models the linear influence of individual characteristics, including an intercept, on the outcome, $\mathbf{x}^{\text{Fe}} =[1,\mathbf{x}]$.
    \item Individual Random Effects: To account for the correlation between repeated measurements from the same individual over time, individual random effects are incorporated into the model. The effects of these random components, $\mathbf{x}^{\text{Re}}$, are modeled using a time-dependent B-spline basis matrix of degree 2.
    \item Interaction with latent Cluster: This is included to model the impact of exposure variables, as represented by their latent clusters, and their interaction with an individual characteristics $\mathbf{x}^{\text{Int}} = [1,\mathbf{x}_{1}]$.
\end{itemize}

This modeling framework allows us to realistically capture both the effects of individual characteristics and exposure variables, as well as their potential interactions, while accounting for the correlated structure of repeated measurements from the same individual.\\

To the best of our knowledge, no existing method enables the direct analysis of such complex data, hindering a rigorous comparison.
Nevertheless, we will benchmark our methodology against the following two-step approaches, which incorporate increasing levels of prior knowledge regarding the latent cluster structure:
\begin{enumerate}
    \item \textbf{True Centroids and Variance\label{meth:TCen}}: This method uses the true centroids and variance parameters of the latent clusters to define the cluster membership used in the regression.
    \item \textbf{True Cluster Assignment\label{meth:Tclus}}: This method represents the ideal scenario, where the true cluster assignments are directly incorporated into the regression analysis.
\end{enumerate}

\begin{figure}
    \centering
    \includegraphics[width=1\linewidth]{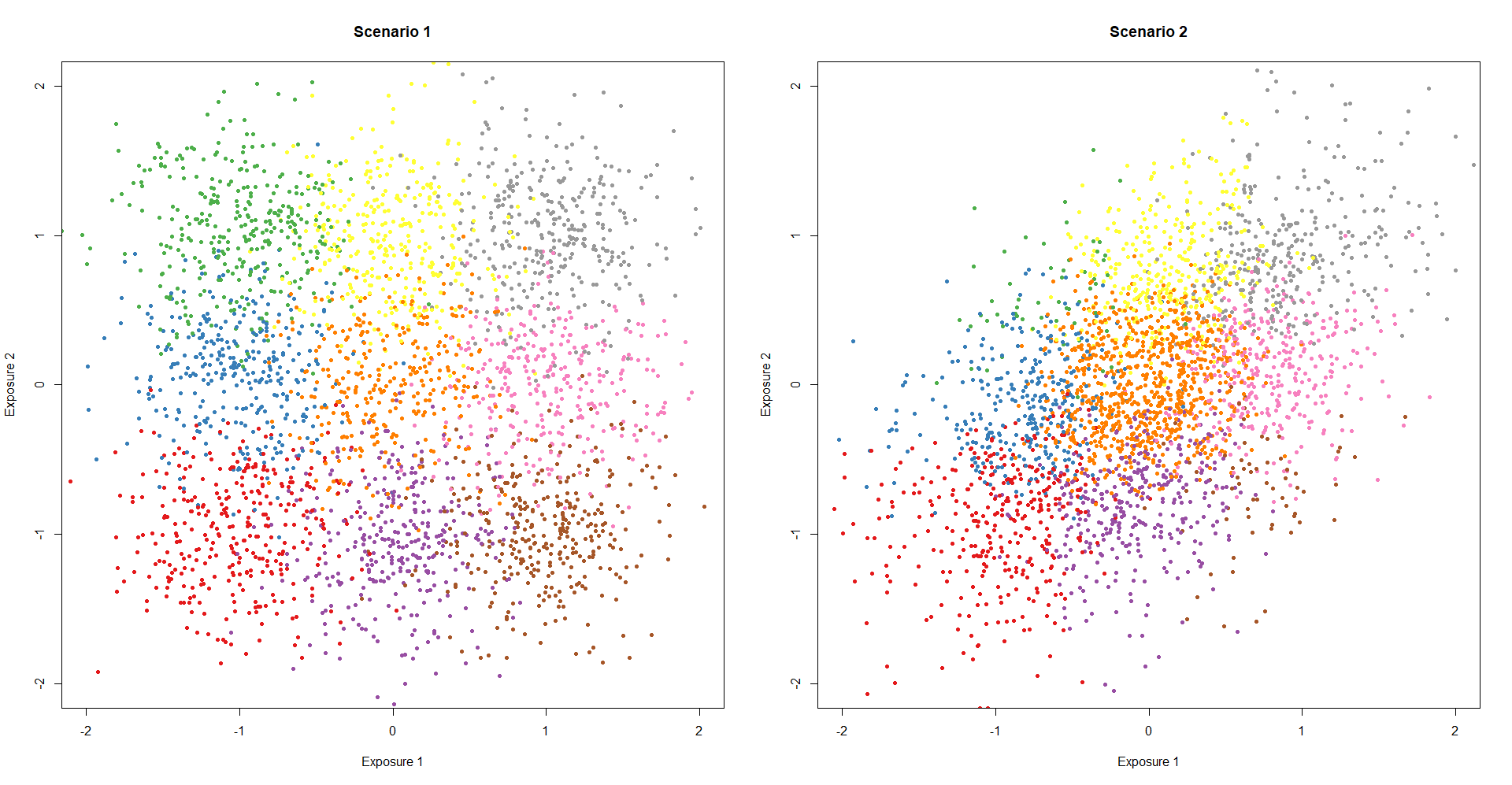}
    \caption{Distribution of observed exposure data for each scenario, with colors indicating latent cluster assignments. In Scenario 1 (left panel), the observed data distribution closely aligns with the underlying latent cluster structure. In Scenario 2 (right panel), the observed exposures are correlated. Both scenarios share the same cluster structure.}
    \label{fig:expdist}
\end{figure}

Furthermore, we will evaluate all these methodologies across two distinct scenarios. In the simpler scenario (Scenario 1), the nine latent exposure clusters are distinctly discernible within the observed exposure distribution, as illustrated in the left panel of Figure \ref{fig:expdist}. Scenario 2 more accurately reflects real-world data characteristics, where the observed exposure distribution exhibits substantial deviation from the underlying latent cluster structure. Specifically, the observed exposures in Scenario 2 demonstrate a strong correlation and sparse representation in the upper-left and lower-right quadrants of the exposure plane (right panel of Figure \ref{fig:expdist}). This sparsity reflects the infrequent occurrence of locations characterized by high levels of one exposure and low levels of the other. Essentially, Scenario 2 significantly increases the informational burden on the LMM for inferring the latent classes.

The comparison of our method against these benchmark approaches and across these two contrasting data scenarios will effectively demonstrate the robustness and power of our proposed methodology.

\subsection{Results}

This section is structured into two main parts. The first part presents an analysis of the model's general performance, with a focus on two primary aspects: the accuracy of LMM parameter inference and the precision of exposure clustering. These two evaluations are performed over 25 independent repetitions of the experiment.\\
The second part provides an illustration, on a specific model fit, of the estimated parameters of the latent clusters. This highlights our model's ability to estimate both the exposure's latent cluster structure as well as their effect on the outcome.

\subsubsection{General performance}

Clustering accuracy is assessed using two distinct metrics. Firstly, consistent with established practices in the Bayesian clustering literature \citep{dahl_multiple_2007,medvedovic_bayesian_2004,fritsch_improved_2009}, we use the adjusted Rand index (ARI) to compare the inferred clusters against the true cluster assignments. The unadjusted Rand index, a measure of similarity between two partitions ranging from 0 (no similarity) to 1 (identical structures), is susceptible to spurious agreements, particularly when cluster sizes or numbers vary.  ARI mitigates this limitation by correcting for chance agreement, thus providing a more robust measure of clustering quality \citep{hubert_comparing_1985}. The Adjusted Rand Index is computed using the \texttt{mclust} R package \citep{scrucca_mclust_2016}. Secondly, we employ the Purity score, defined as: for each cluster, the number of data points belonging to the most frequent true class within that cluster, summed across all clusters, and then divided by the total number of data points. This metric ranges from 0 to 1, with 1 indicating perfect purity. A downside of the Purity score is that it does not penalize an increase in the number of estimated clusters. However, when the number of estimated clusters is equal to or less than the true number of clusters, this score can be directly interpreted as the proportion of correctly classified data points.

\begin{figure}[h!]
    \centering
    \includegraphics[width=1\linewidth]{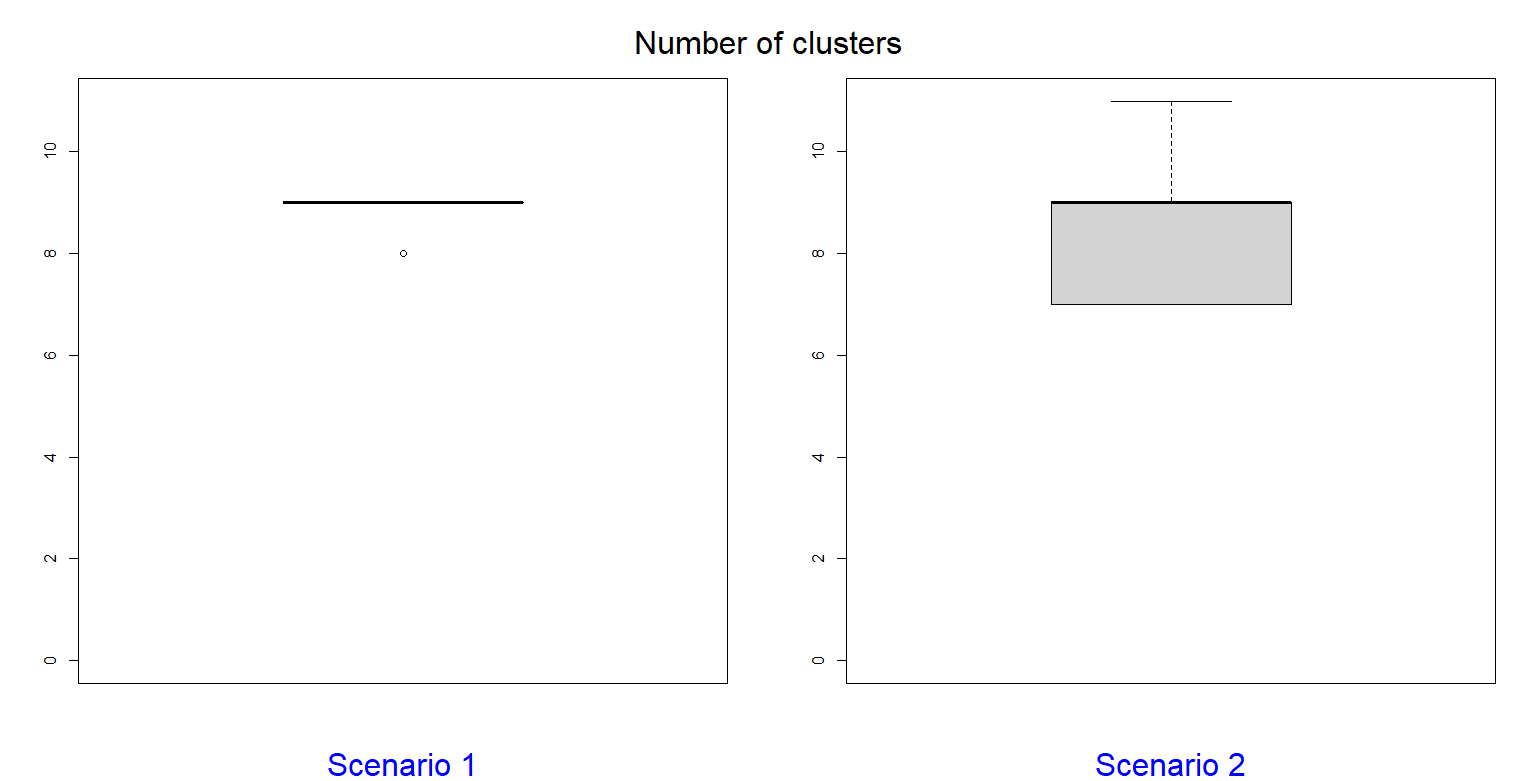}
    \caption{Number of clusters of the Profile LMM representative clustering. Distribution over 25 experiments. }
    \label{fig:nclus}
\end{figure}

\begin{figure}[h!]
    \centering
    \includegraphics[width=1\linewidth]{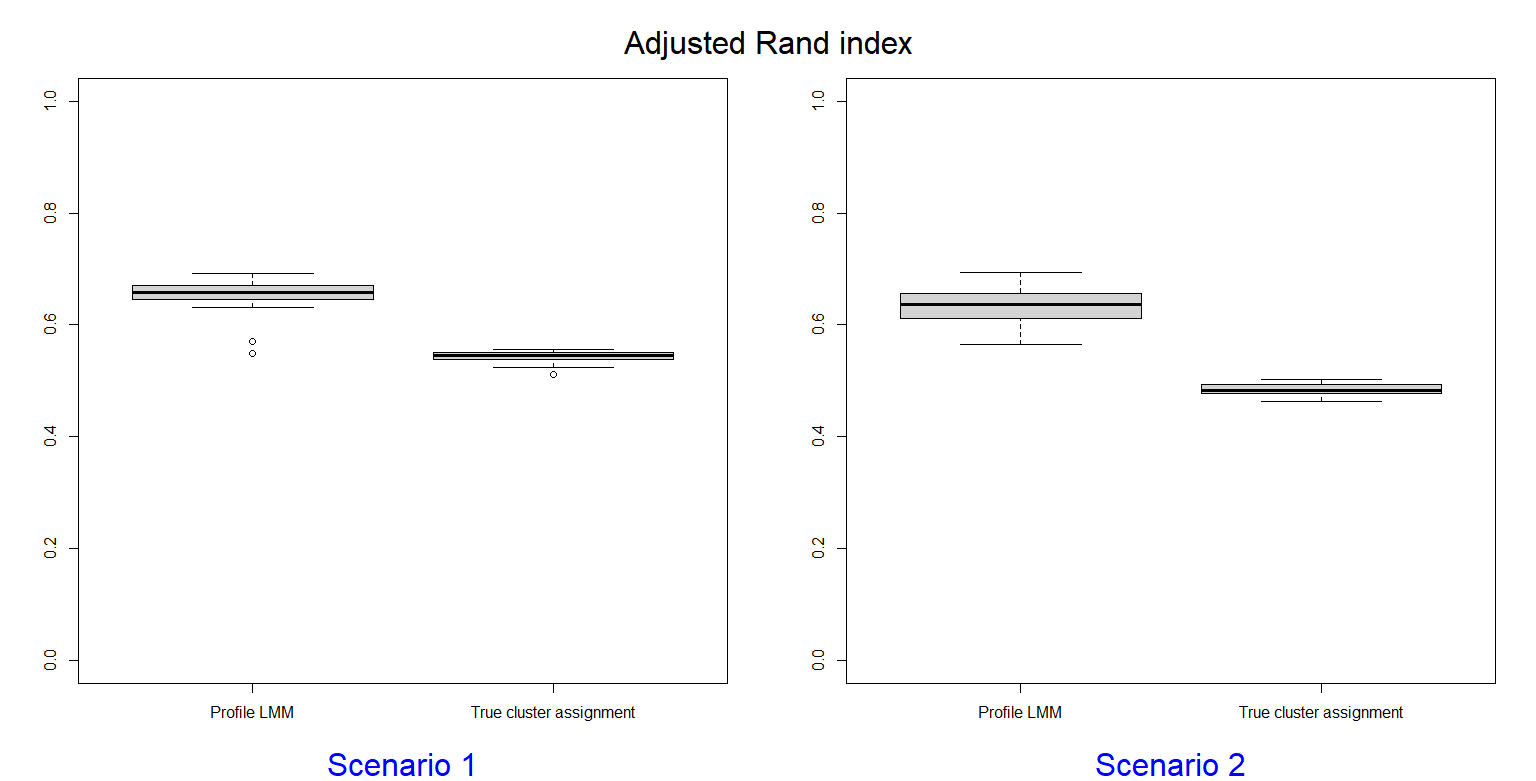}
    \caption{Adjusted Rand Index. Distribution over 25 experiments. }
    \label{fig:Pear}
\end{figure}

\begin{figure}[h!]
    \centering
    \includegraphics[width=1\linewidth]{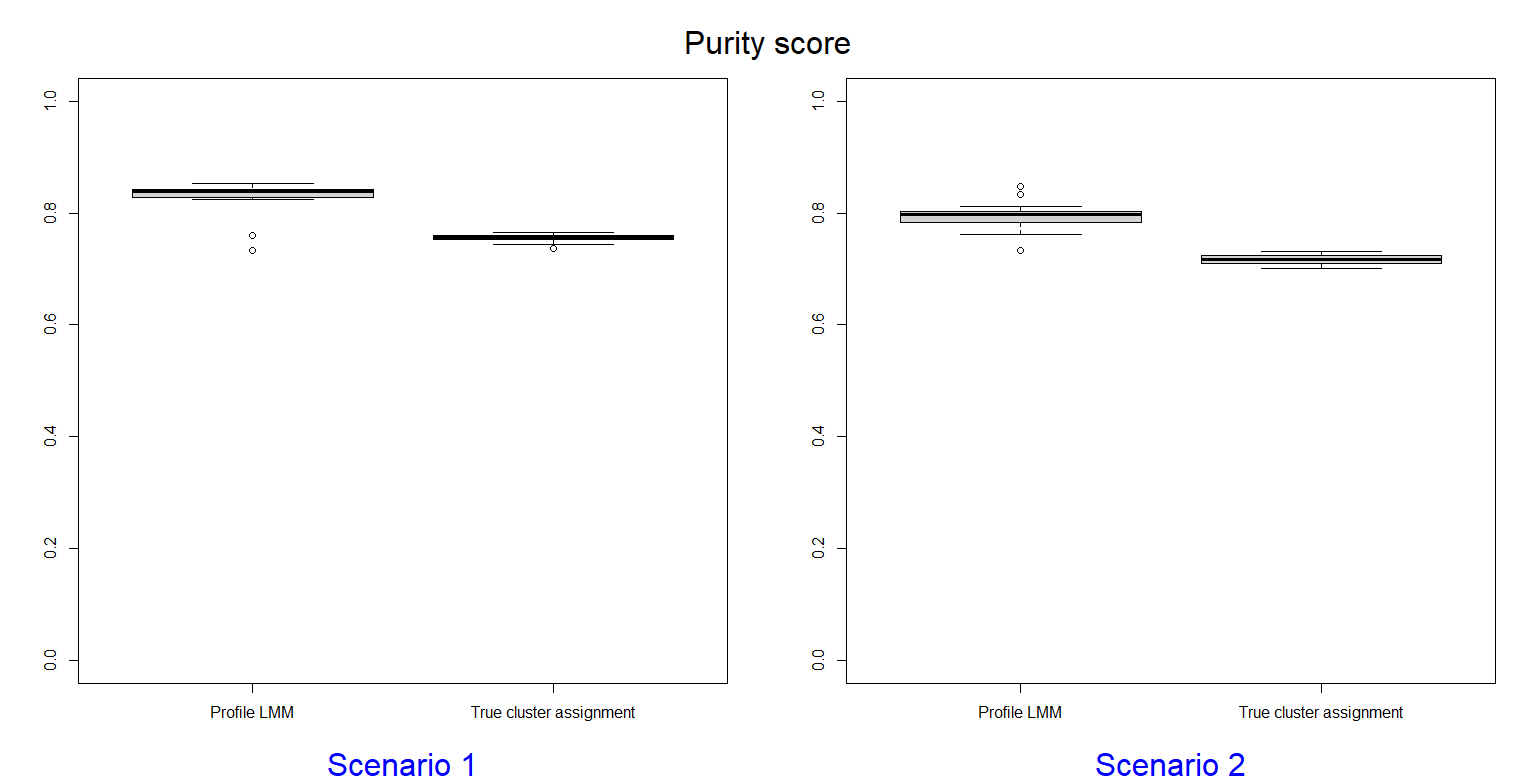}
    \caption{Purity score. Distribution over 25 experiments. }
    \label{fig:Purity}
\end{figure}

Supplementary Figure \ref{fig:nclus} presents the number of clusters identified by the Profile LMM's representative clustering. Supplementary Figures \ref{fig:Pear} and \ref{fig:Purity} subsequently compare the ARI and Purity score, respectively, for the Profile LMM's representative clustering against those obtained from GMM clusters derived using the true latent centroids and covariances.

Our findings indicate that Profile LMM consistently outperforms the GMM using the true latent centroids and covariances in both evaluation scores and in both scenarios. This superior performance demonstrates Profile LMM's capability to accurately identify the underlying latent cluster structure and effectively leverage the mixed model results for improved classification of data points situated between clusters. Furthermore, we observed that increasing the complexity of the scenario — and thus augmenting the information that the profile-LMM must extract from its LMM component — marginally impacts the model's ability to discern the correct latent cluster structure. 

For the evaluation of the LMM parameter inference, we report the relative Root Mean Square Error (RMSE) of the parameters, calculated as $||\hat{\theta}-\theta_0||/||\theta_0||$. Here, the norm represents the Euclidean norm for vector parameters and the Frobenius norm for matrix parameters. We partitioned the fixed effects coefficients into two groups: those that interact with the exposure latent clusters (the intercept and $\mathbf{x}_1$) and those who do not ($\mathbf{x}_2$,$\mathbf{x}_3$,$\mathbf{x}_4$).

\begin{figure}
    \centering
    \includegraphics[width=1\linewidth]{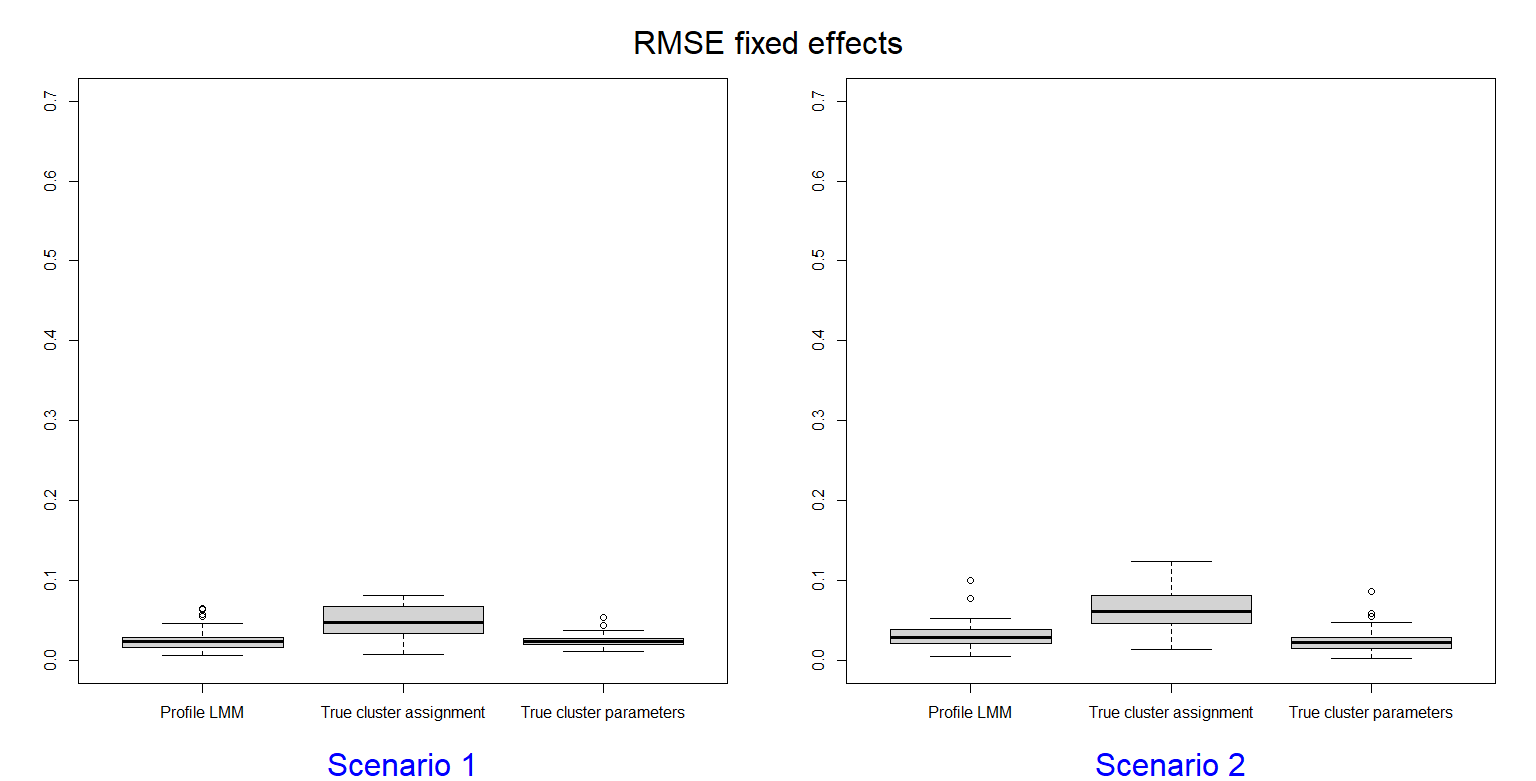}
    \caption{Relative-RMSE of the fixed effect estimates of the covariates ($\mathbf{x}_{2},\mathbf{x}_{3},\mathbf{x}_{4}$) that do not interact with the latent exposure cluster. Distribution over 25 experiments.}
    \label{fig:RMSEFE}
\end{figure}

\begin{figure}[h!]
    \centering
    \includegraphics[width=1\linewidth]{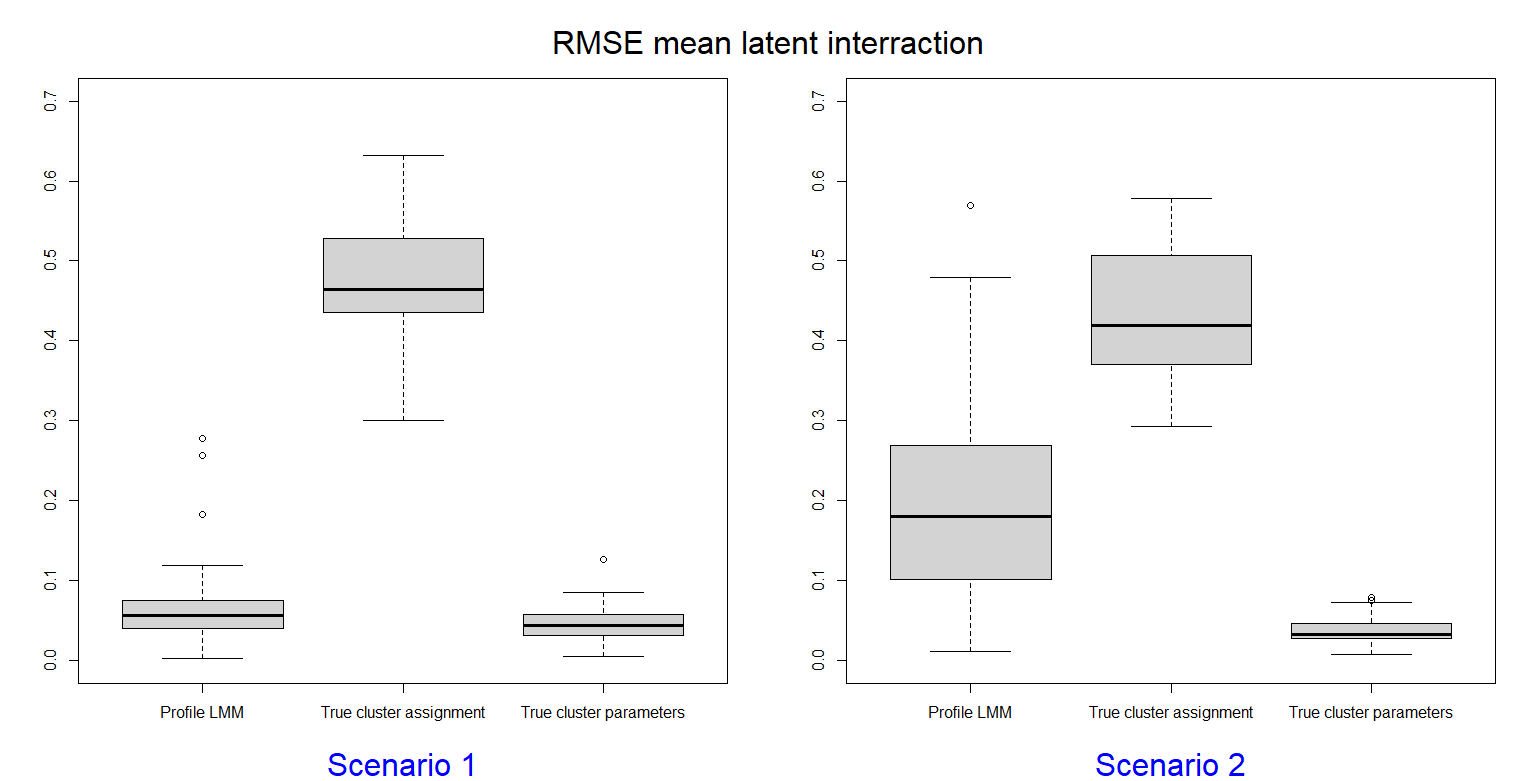}
    \caption{Relative-RMSE of the fixed effect estimates of the covariates (intercept and $\mathbf{x}_{1}$) that interact with the latent exposure cluster. Distribution over 25 experiments.}
    \label{fig:RMSEFEINT}
\end{figure}

\begin{figure}[h!]
    \centering
    \includegraphics[width=1\linewidth]{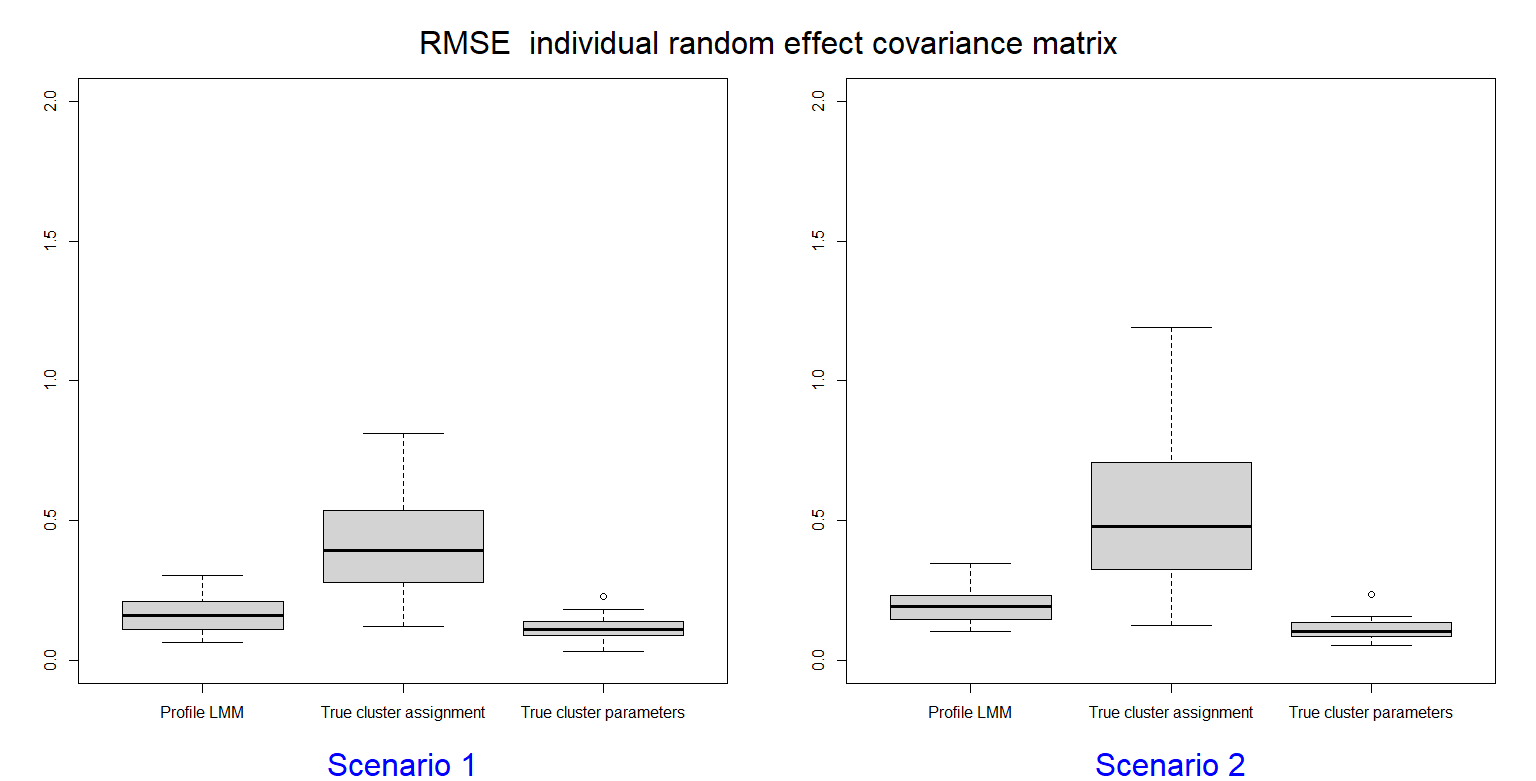}
    \caption{Relative-RMSE of the individual random effect covariance matrix estimates. Distribution over 25 experiments.}
    \label{fig:RMSERE}
\end{figure}

Supplementary Figure \ref{fig:RMSEFE} presents boxplots illustrating the relative-RMSE for the fixed effect parameters of the first group, while Supplementary Figure \ref{fig:RMSEFEINT} shows the results for the second. Supplementary Figure \ref{fig:RMSERE} displays the relative-RMSE for the covariance of the individual random effects. Our observations indicate that the coefficients that interact with the latent clusters of exposure are the most affected by the increased complexity of the model. However, our model's performance remains superior to those derived using the true latent centroids. For the fixed effects that do not interact with the exposures, the coefficients appear stable across both scenarios for the two methods, with the gain from using the profile-LMM being limited.
Overall, and consistent with the clustering scores, the profile-LMM demonstrates superior performance in parameter inference across both scenarios.

A direct evaluation of the inference of the latent cluster effect on the outcome, $\gamma$, is not presented here due to its strong dependence on cluster assignment, which complicates comparisons between experiments. However, we posit that accurate cluster retrieval, coupled with precise inference of fixed and random effect parameters, serves as a robust indicator of the accuracy in estimating the latent cluster effects on the outcome. This assertion will be further substantiated in the next section.

\subsection{Cluster characteristics estimation}
In this section, we focus on a single experiment to delve deeper into the characteristics of the estimated clusters. These characteristics comprise the estimated centroids and variance components of the exposure GMM, which provide a comprehensive characterization of the exposure clusters. Subsequently, we examine the estimated effects of each cluster on the outcome variable.

\begin{figure} [ht]
    \centering
    \includegraphics[width=1\linewidth]{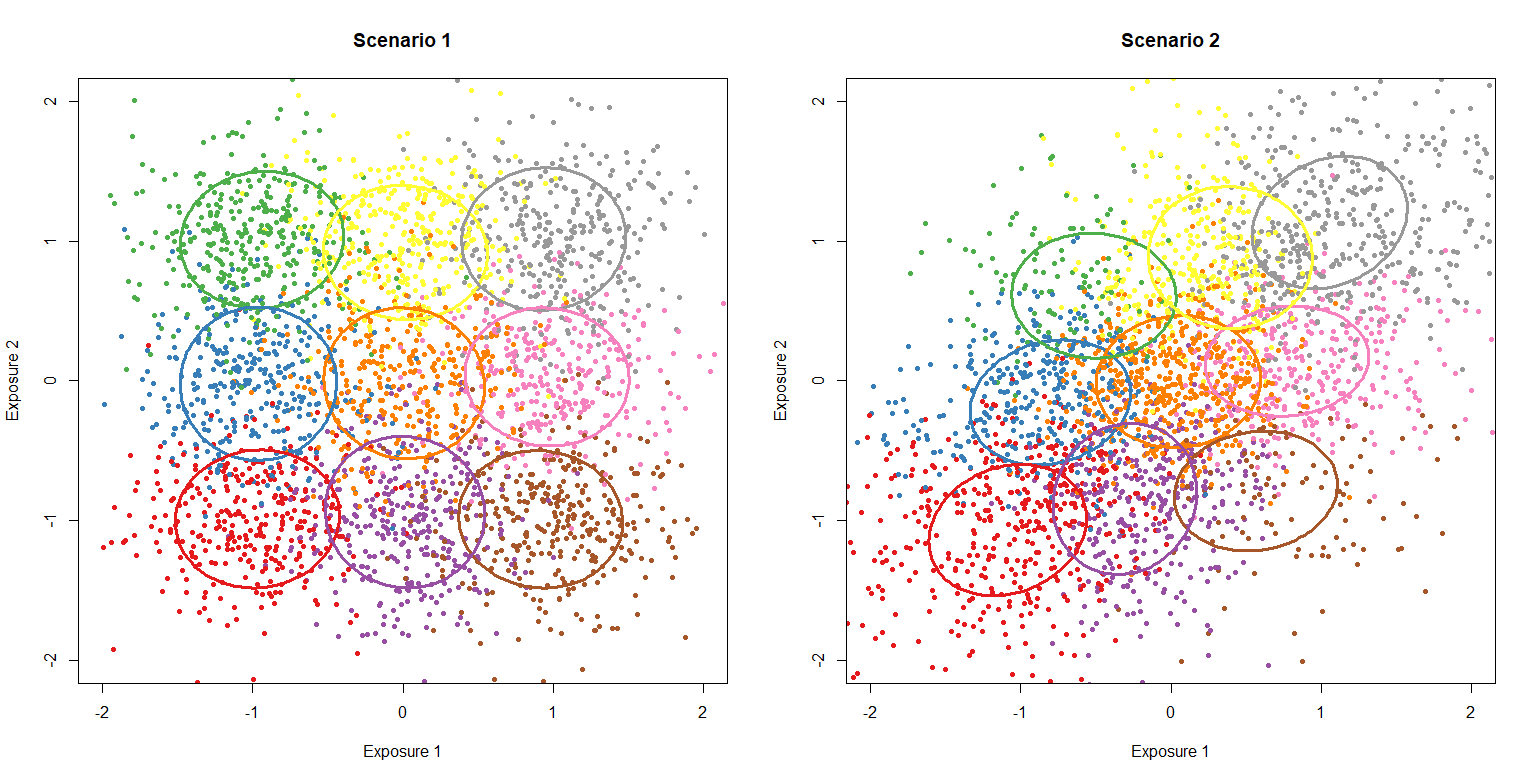}
    \caption{Profile LMM cluster centroids and variance estimates. Points represent observation and their colors match the true latent clusters. Colors of the estimated clusters and true clusters have been matched for readability.}
    \label{fig:clusterGMM}
\end{figure}

\begin{figure}[ht!]
    \centering
    \includegraphics[width=1\linewidth]{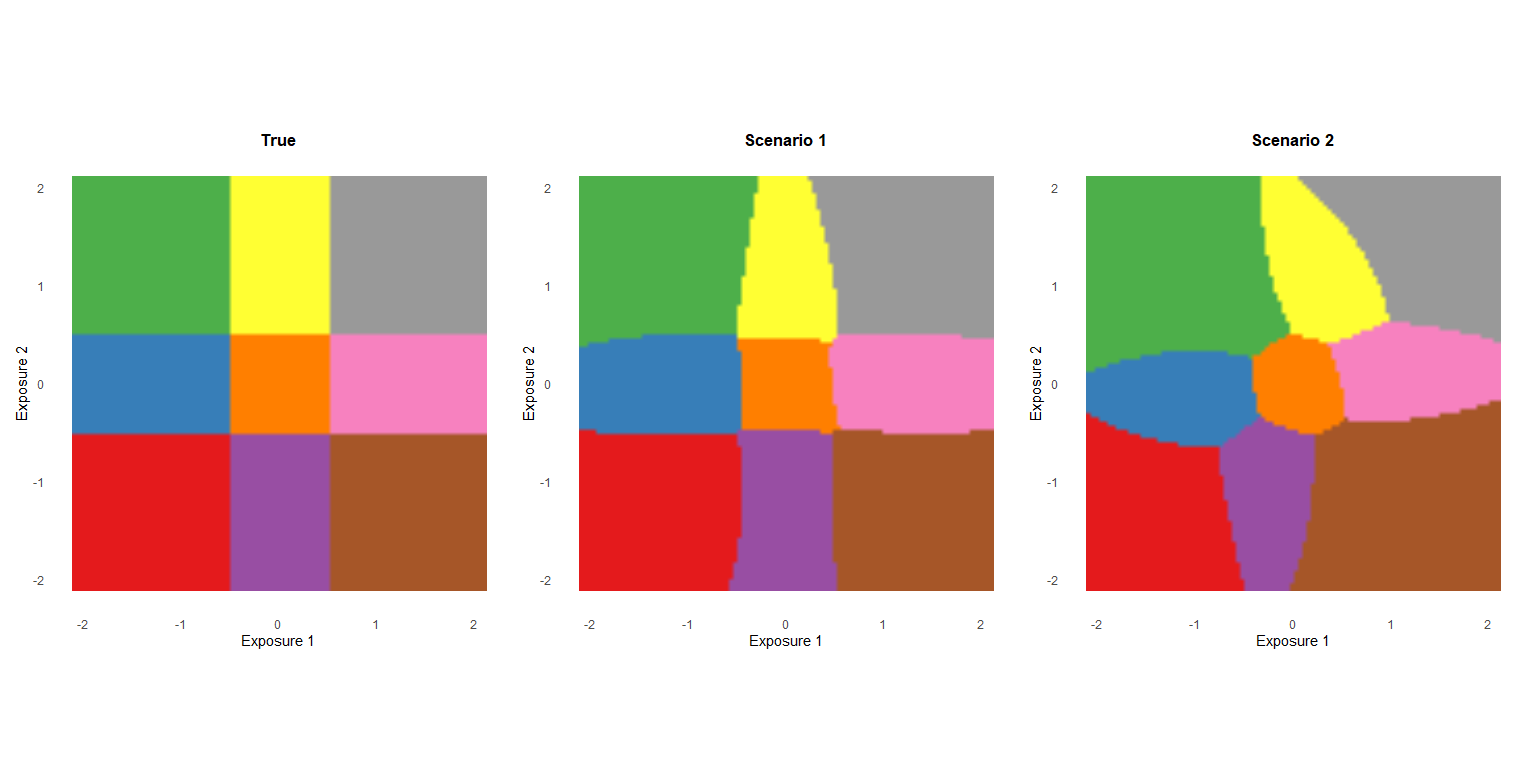}
    \caption{Profile LMM cluster predictive zones. Colors indicate the cluster with the local maximum density.}
    \label{fig:clusterZone}
\end{figure}

Figure \ref{fig:clusterGMM} illustrates the characteristics of the Profile LMM's representative clustering, where each ellipse visually represents a level set of the Gaussian mixture distribution associated with a specific cluster. For both scenarios, the true latent cluster assignment distribution remains identical. In Scenario 2, the correlation of the observed data points slightly influences the placement of the estimated centroids. Supplementary Figure \ref{fig:clusterZone} displays the resulting predictive zones for these clusters, where the color indicates the mixture component with the highest density. Here, too, we note that in Scenario 2, the predictive zones exhibit slight deformations, although our method successfully estimates the general geometry of the mixture model.

Furthermore, Tables \ref{tab:intercept} and \ref{tab:slope} demonstrate that, in addition to accurately retrieving the cluster structure, Profile LMM also precisely estimates the intercept and slope parameters for each of the latent clusters within the outcome LMM.\\

\begin{table}[h!]
\centering
\begin{tabular}{|c|ccccccccc|}
  \hline
 Cluster centroid& -1 -1 & -1 0 & -1 1 & 0 -1 & 0 0 & 0 1 & 1 -1 & 1 0 & 1 1 \\ 
  \hline
True & -4.00 & -1.00 & 2.00 & -3.00 & 0.00 & 3.00 & -2.00 & 1.00 & 4.00 \\ 
  Scenario 1 & -3.98 & -1.12 & 1.94 & -2.88 & -0.07 & 2.68 & -2.11 & 0.91 & 3.94 \\ 
  Scenario 2 & -3.79 & -1.00 & 1.69 & -2.83 & 0.12 & 3.19 & -1.79 & 1.12 & 4.13 \\ 
   \hline
\end{tabular}
\caption{Example of the estimated latent cluster intercepts} 
\label{tab:intercept}
\end{table}
\begin{table}[ht]
\centering
\begin{tabular}{|c|ccccccccc|}
  \hline
Cluster centroid & -1 -1 & -1 0 & -1 1 & 0 -1 & 0 0 & 0 1 & 1 -1 & 1 0 & 1 1 \\ 
  \hline
True & -1.67 & 1.60 & 0.45 & 0.05 & -2.56 & 1.19 & 0.77 & 0.06 & 0.13 \\ 
  Scenario 1 & -1.60 & 1.49 & 0.43 & 0.04 & -2.00 & 0.79 & 0.71 & 0.00 & 0.16 \\ 
  Scenario 2 & -1.54 & 1.41 & 0.67 & -0.23 & -1.95 & 0.75 & 0.75 & 0.16 & 0.23 \\ 
   \hline
\end{tabular}
\caption{Example of the estimated latent cluster slopes} 
\label{tab:slope}
\end{table}

Collectively, these simulation results indicate that Profile LMM effectively leverages the outcome regression to infer the latent cluster structure. Moreover, the shared information between the LMM and clustering models leads to superior point clustering performance compared to approaches that rely on a-priori knowledge of the true latent clusters. In turn, this improved clustering accuracy facilitates a more accurate estimation of the LMM parameters.

\section{Application to diastolic blood pressure and exposure estimation}\label{sec:RealData}

Our methodology is applied to the analysis of real-world exposure data, specifically focusing on measuring the impact of multiple environmental pollutants and green space density on DBP. This problem presents a compelling application for our method. As highlighted by \cite{coker_multi-pollutant_2018}, inter-pollutant correlations (illustrated in Supplementary Figure \ref{tab:corExp}) can affect effect estimates when conventional regression techniques, if multiple exposures are added to the model. Furthermore, while recent advancements in the study of pollutant mixtures \citep{oakes_evaluating_2014} show promise, these approaches often rely on predefined mixtures, underscoring the absence of a unified analytical framework.

Through this application, we demonstrate the capacity of our method to handle such problems. In addition to accurately estimating the effects of standard confounding variables, our method identifies distinct exposure clusters and their interactions with individual-level covariates.

\subsection{Data and model specification}
We utilize the Lifelines cohort dataset \citep{sijtsma_cohort_2022}, a longitudinal dataset tracking over 160,000 individuals for more than 15 years. The project is currently collecting its third wave, and the data set comprises over 323,000 observations unevenly distributed across waves: 162,000 individuals in the first wave, 118,000 in the second, and 41,000 in the third. Individual characteristics are available for each wave. In addition to the information provided by the Lifelines dataset, we enrich the dataset with environmental exposure data provided by the GECCO platform \citep{gecco_consortium_deep_2020}. These exposures are assessed at the geocoded home addresses of the participants.

We now specify the details of our model, where DBP serves as our outcome variable. We incorporate a set of commonly recognized confounders as fixed effects: sex, age, body mass index (BMI), marital status, ethnicity, birthplace, current employment status, highest educational degree attained and current smoker status. Given the longitudinal nature of our data, a natural cubic spline basis of degree 3 is included as a random effect at the individual level to model the correlation among repeated observations. Finally, for clustering covariates, we use annual average outdoor pollution concentrations at participants' home addresses for nitrogen dioxide (NO$_2$), ozone (OZO) and particulate matter with diameters less than 2.5 $\mu$m (PM25). Additionally, green space density, assessed within a 0.3 km buffer zone around the home address (NDV\_MD3), is included. Within this model, we evaluate the effect of each latent cluster on DBP and investigate their interaction with gender, age, and current smoker status.

To verify the stability of our results and for computational reasons, the dataset is partitioned into two non-overlapping subsets of 60,000 observations. This separation is performed at the individual level, ensuring that all observations pertaining to a single individual reside within the same dataset. Our profile regression estimation is performed by simulating an MCMC chain of 15,000 iterations, with the initial 5,000 iterations discarded as burn-in. In this process a maximum of 60 clusters are considered. Model convergence is verified by examining the trace plot of the Dirichlet process concentration parameter (see Supplementary Figure \ref{fig:conc}).

\subsection{Results}

\begin{figure} [ht]
    \centering
    \includegraphics[width=1\linewidth]{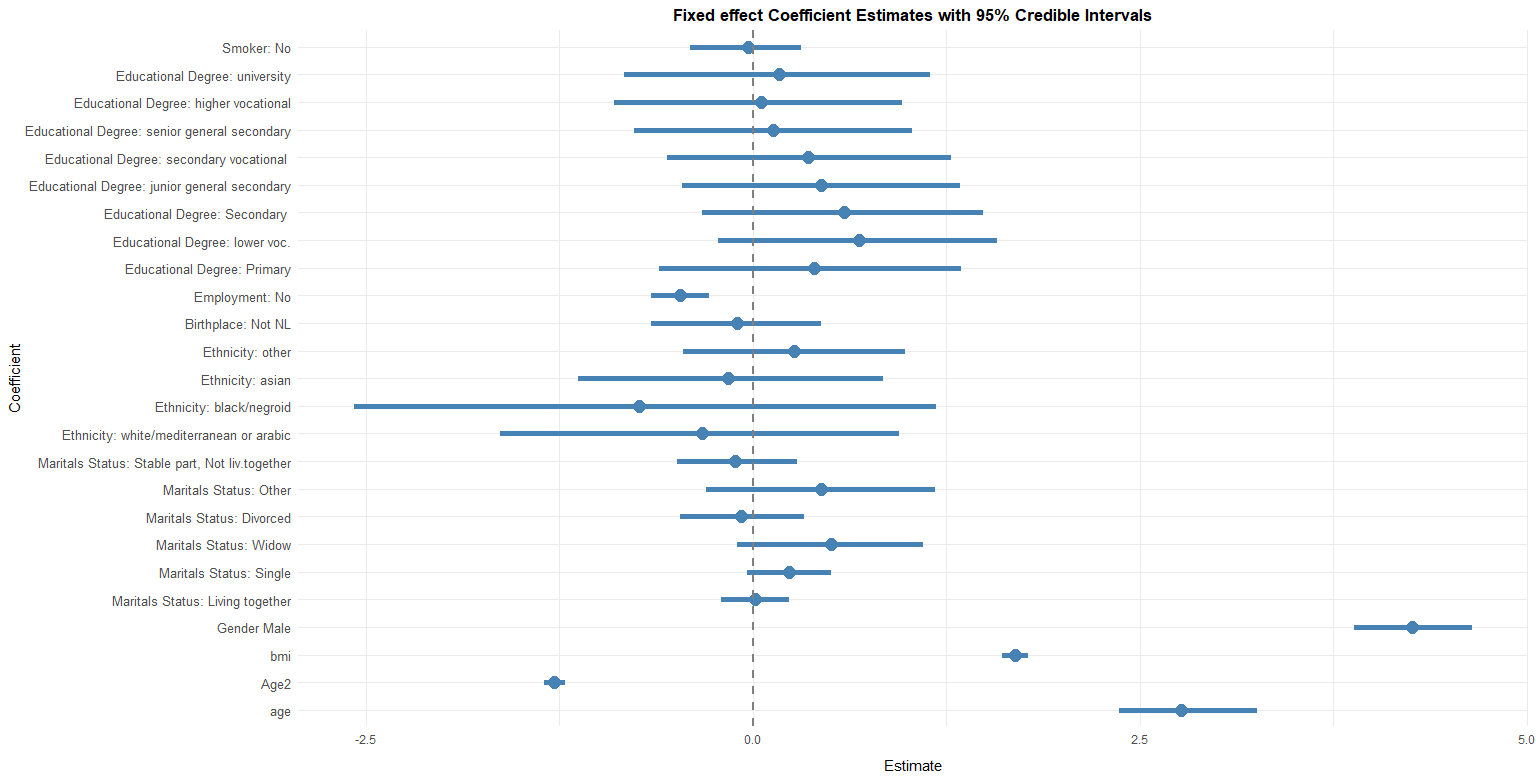}
    \caption{Fixed effect estimates with 95\% credible intervals of standardized coefficients, DBP as an outcome.}
    \label{fig:FEReal}
\end{figure}

We begin our analysis by an examination of the fixed effect estimates presented in Figure \ref{fig:FEReal} with their corresponding 95\% credible intervals. The fixed effect estimates indicate that the model successfully identifies well-established determinants of DBP. Specifically, the model captures the significant effects of factors such as gender and BMI, in addition to a non-linear effect of age \citep{reckelhoff_gender_2001,neter_influence_2003,vishram_impact_2012}. This nonlinear relationship is characterized by an initial increase in DBP that peaks around age 50-60 and can then stabilize or even fall at older age due to arterial stiffening. In our model, this phenomenon is captured by including a positive linear coefficient for age and a negative coefficient for age squared (\texttt{Age2}). \\

We then turn our attention to the estimation of the cluster mixtures. To estimate the cluster mixtures, the observation similarity matrix was constructed using $10,000$ observations. The co-occurrence counts for this matrix were averaged over $8,000$ simulations from a converged MCMC chain. Although we allowed for the potential discovery of up to $30$ clusters, the PAM algorithm ultimately identified five distinct exposure clusters from the similarity matrix. The distribution of observations in these clusters is detailed in Table \ref{tab:Clus}. Due to to its sparse representation, Cluster 5 was subsequently excluded from all further analysis. 

\begin{figure} [ht]
    \centering
    \includegraphics[width=1\linewidth]{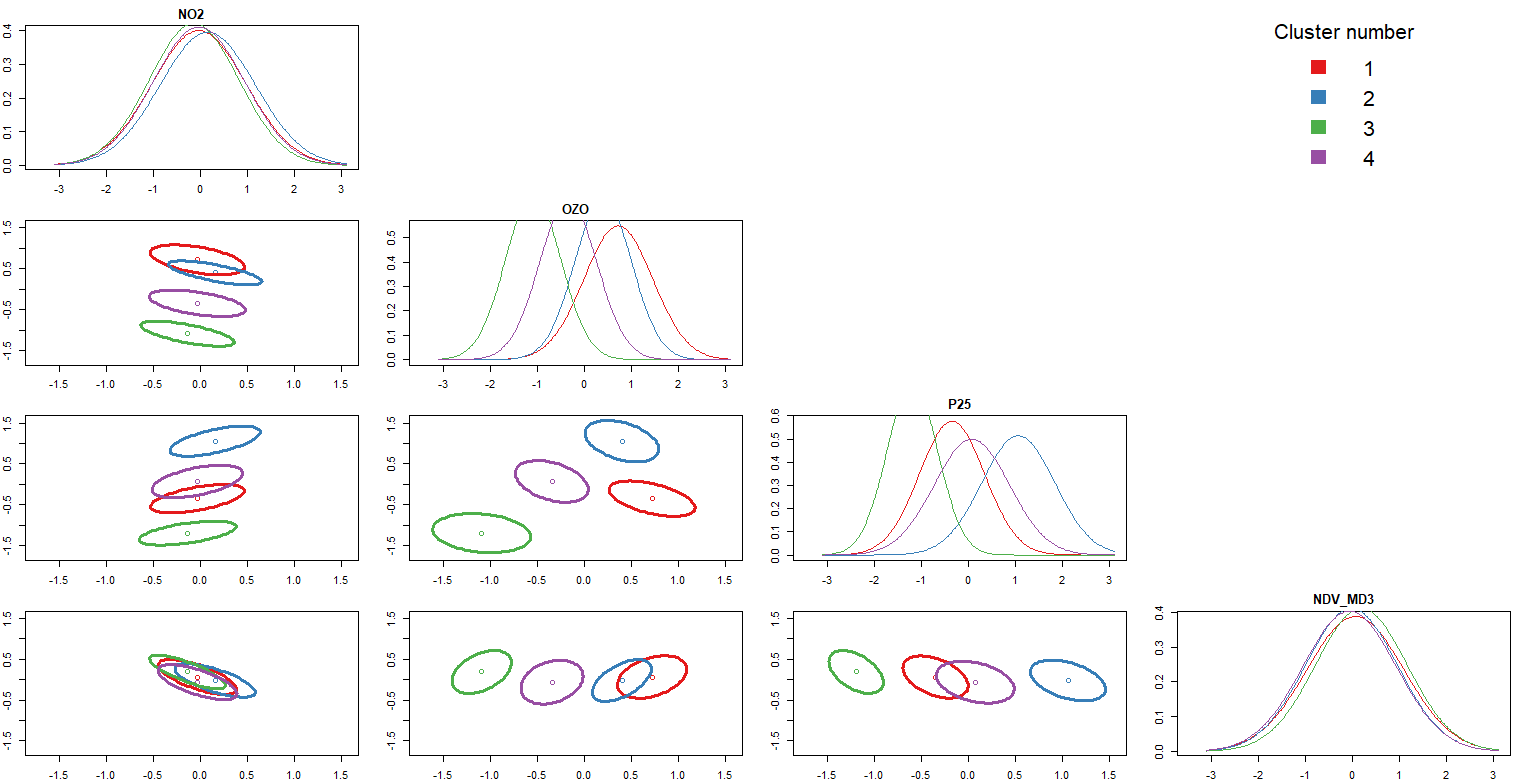}
    \caption{Exposure latent clusters mixtures representations. Diagonal panels show the marginals for each exposure and off-diagonal panels the pairwise marginal distributions. The dot indicates the centroid of the mixture and the ellipse is a representation of the covariance matrix. }
    \label{fig:clusterReal}
\end{figure}

\begin{table}[ht]
\centering
\begin{tabular}{|c|ccccc|}
  \hline
Cluster  & 1 & 2 & 3 & 4 & 5  \\ 
  \hline
Nb. observations & 2697 & 1103 & 4542 & 1618 & 50 \\ 
   \hline
\end{tabular}
\caption{Distribution of the 10,000 observations among the 5 clusters identified by the profile LMM. Cluster 5 is removed from the rest of the analysis, given the low number of observation it contains.} 
\label{tab:Clus}
\end{table}
Figure \ref{fig:clusterReal} visualizes the marginal distributions of the Gaussian mixture components of the GMM for each identified cluster, with a unique color assigned to each. The diagonal panels of this figure display the univariate marginal distributions, while the off-diagonal panels illustrate the pairwise bivariate marginal distributions. In each panel, the cluster mean is represented by a central point, and the covariance structure is represented by an encompassing ellipse. The ellipses, which are scaled to share a common radius, depict the correlation among components: a more circular ellipse indicates a higher degree of independence among the components within that cluster.

Analysis of the identified clusters reveals the most pronounced distinctions in ozone and PM25 exposures. Specifically, Cluster 1 comprises observations characterized by elevated ozone concentrations and comparatively lower PM25 exposures. Conversely, Cluster 2 exhibits moderate ozone levels coupled with high PM25 exposures. Cluster 3 is distinguished by low exposures to both pollutants, while Cluster 4 demonstrates average exposures for both. Minimal inter-cluster differences were observed for NO$_2$ and NDV\_MD3. \\

\begin{figure}[ht] 
    \centering
    \includegraphics[width=1\linewidth]{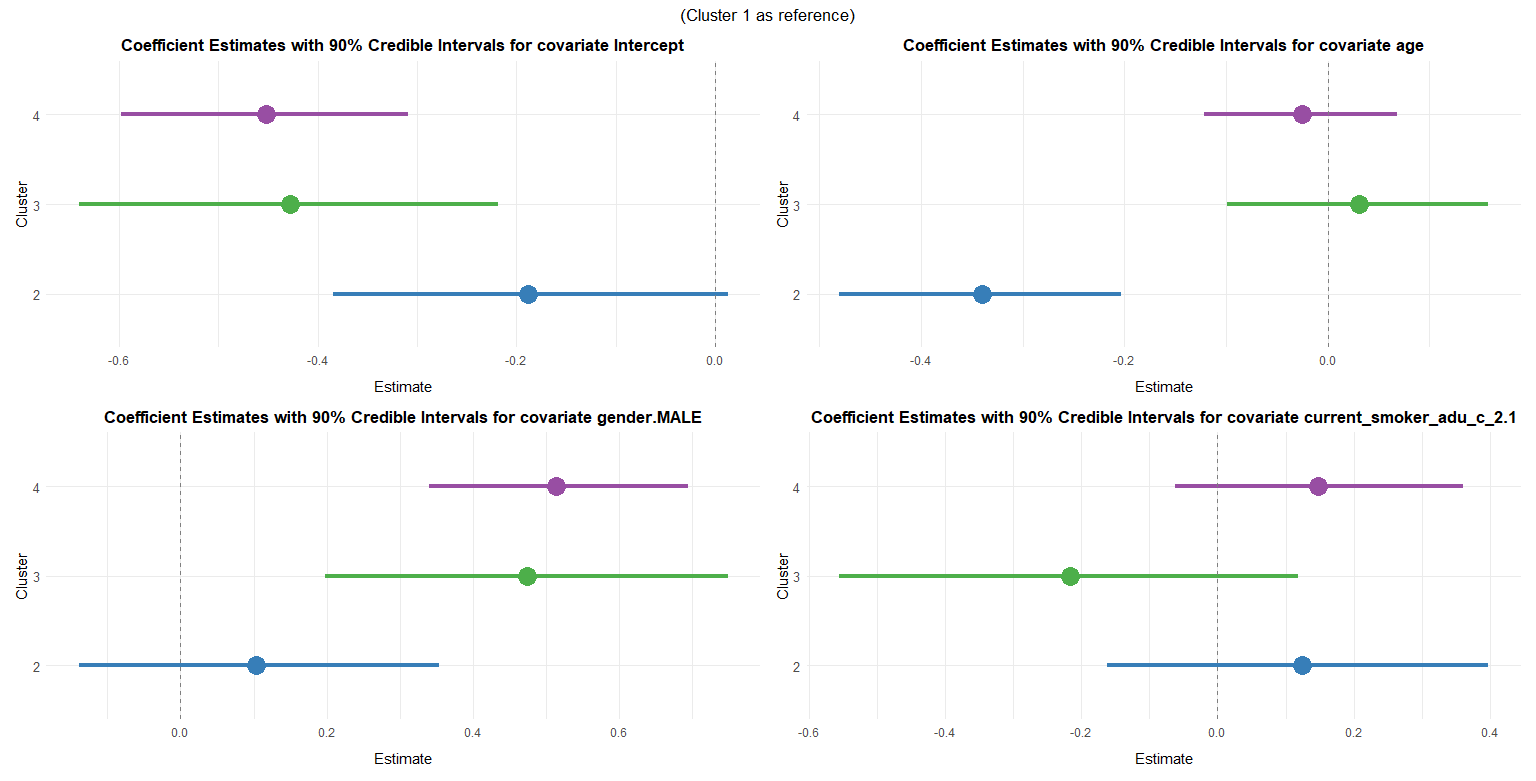}
    \caption{Exposure latent clusters effect on the outcome using cluster 1 as reference. The cluster number (on the y-axis) and the color match the ones of Figure \ref{fig:clusterReal}. }
    \label{fig:clusterRealEffect}
\end{figure}

Figure \ref{fig:clusterRealEffect} illustrates the estimated cluster effects, with the last cluster serving as the reference for comparison. The reported values represent the difference in effect relative to this reference, accompanied by their respective 90\% credible intervals.
A notable finding is the significant divergence of Clusters 3 and 4 from Cluster 1, in terms of both the intercept and the interaction with gender. This suggests that varying ozone exposure levels affect differently males and females; specifically, elevated ozone appears to increase DBP in males while decreasing it in females (negative intercepts for both Cluster 3 and 4 in Figure  \ref{fig:clusterRealEffect}).
Furthermore, a diminished influence of age on DBP is observed in Clusters 2 when compared to Clusters 1, 3 and 4. This pattern suggests that PM25 exposure may modulate the age-related trajectory of blood pressure.
Conversely, the impact of smoking on DBP does not appear to be significantly influenced by the various pollutant exposure clusters. \\

\begin{figure}[h!]
    \centering
    \includegraphics[width=1\linewidth]{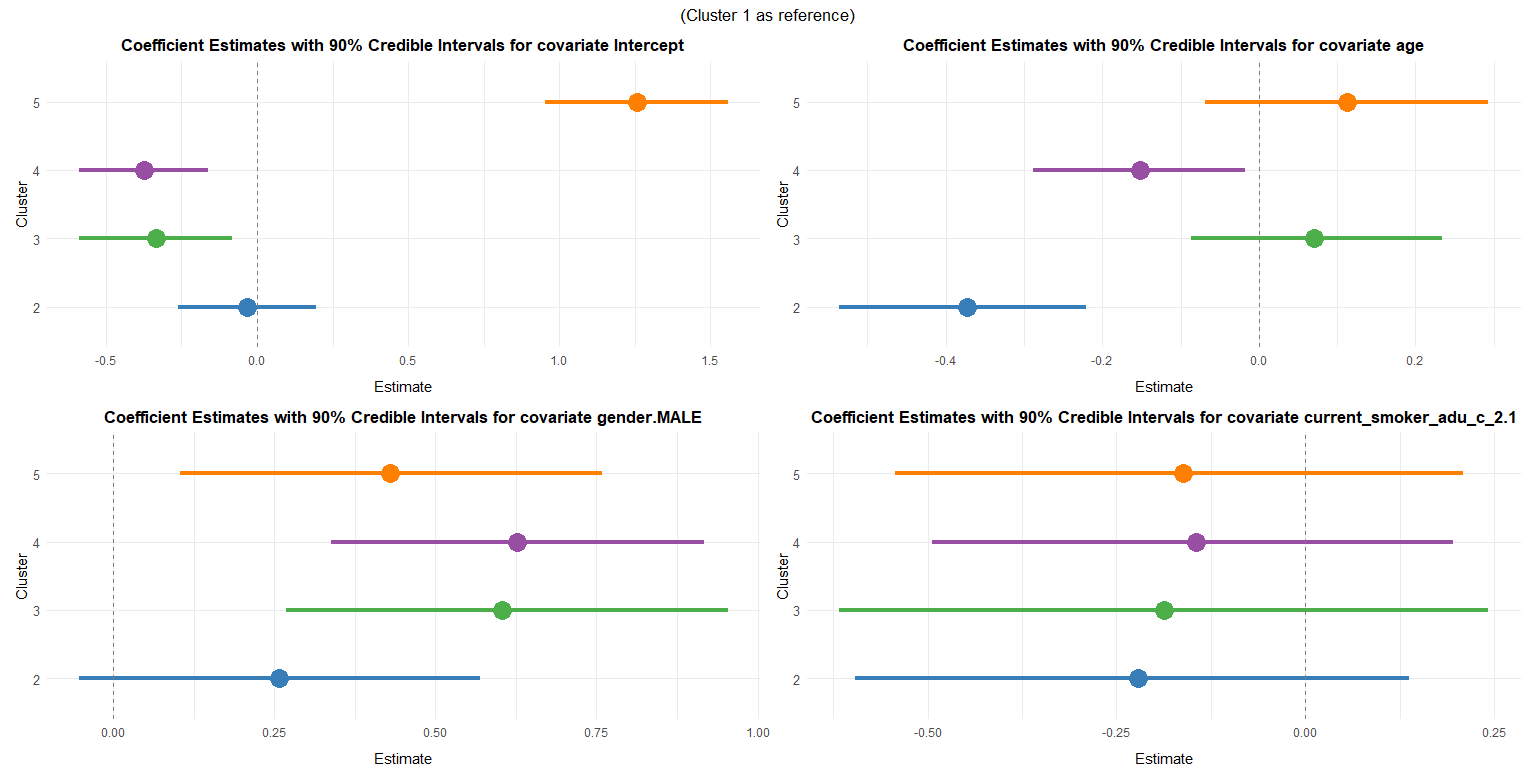}
    \caption{Exposure latent clusters effect on the outcome. Cluster 1 as reference. Estimation done on a different subset of the data. }
    \label{fig:clusterRealEffectStab}
\end{figure}

These findings are further corroborated by the sensitivity analysis presented in the Supplementary Material. This analysis, conducted on a second subset of the data (see Supplementary Figures \ref{fig:FERealStab}, \ref{fig:clusterRealStab}, and \ref{fig:clusterRealEffectStab}), also identifies the four main clusters found in the primary analysis, along with their respective effects. The principal difference is that, in this secondary analysis, Cluster 5 contains a non-negligible proportion of the clustered observations. However, as illustrated in Supplementary Figure \ref{fig:clusterRealStab}, this cluster displays a more uniform distribution across all pollutant exposures. This characteristic suggests that it warrants cautious interpretation and validates the decision to exclude it from the main analysis.

\section{Discussion}

In this paper we introduce a novel profile-LMM framework that effectively addresses the three key challenges identified in the introduction: high collinearity among exposures, the need to account for repeated measures, and the capacity to model interactions with individual-level characteristics. Our method builds on the strengths of both Bayesian profile regression and LMM. Bayesian profile allows to identify an unconstrained number of clusters among the correlated exposures. LMM allows to have a very flexible framework to model the effect of the latent cluster memberships and covariates on the outcomne. Moreover, the fully Bayesian nature of our inference procedure allows to quantify and take into account the estimation uncertainty both in LMM parameters and cluster allocations.\\
The results from both our simulated and real-world analyses demonstrate the practical utility and robustness of this approach. Our simulation study confirmed the method's ability to accurately estimate the parameters of the underlying LMM and, critically, to identify the correct latent cluster structure in increasingly complex setups. The model is, in particular, able to identify the exposure cluster even when the observed exposure distribution is independent to its latent structure. This validation provides a strong foundation for the interpretability of our findings from the Lifelines cohort. The application to the Lifelines dataset successfully identified distinct exposure clusters that are significantly associated with DPB. These clusters represent combinations of pollutants and neighborhood metrics, providing a more comprehensive and plausible understanding of environmental influences on DBP compared to traditional regression methods that treat each exposure in isolation.

While our method represents a significant step forward, it is not without limitations. First, fitting the full Bayesian profile-LMM can be computationally intensive compared to standard models, especially when dealing with very large datasets or a high number of covariates or exposures. The latter limitation could be attenuated by adding a selection procedure for both the regression covariates and exposures. Future work could incorporate sparsity-inducing priors to address this and handle high-dimensional covariate spaces more effectively. Our model is also currently limited to continuous outcomes, but it could be extended to accommodate other data types, such as binary or time to event data, through the use of Generalized Linear Mixed Models. 

In conclusion, the proposed profile-LMM offers a powerful, flexible, and interpretable framework for analyzing complex longitudinal exposome data. It moves beyond the limitations of existing methods to provide a more comprehensive understanding of how diverse environmental factors and their interactions with individual characteristics jointly influence health outcomes. We believe that this methodology will be a valuable tool for future exposome studies. Future research will focus on developing a user-friendly R package to make this approach more accessible to the wider biomedical research community and extending the model to handle other types of outcomes, such as time-to-event or categorical data.

\section*{Acknowledgments}

We thank Haykanush Ohanyan for her assistance with the data preprocessing. \\
Geo-data were collected as part of the Geoscience and Health Cohort Consortium (GECCO), which was financially supported by the Netherlands Organisation for Scientific Research (NWO), the Netherlands Organisation for Health Research and Development (ZonMw), and Amsterdam UMC. More information on GECCO can be found on www.gecco.nl. This work was also supported by EXPOSOME-NL, funded through the Gravitation program of the Dutch Ministry of Education, Culture, and Science and the Netherlands Organization for Scientific Research (NWO grant number 024.004.017).\\
All included studies were approved by an ethical committee. 
Since the data underlying this article contain privacy-sensitive data, access is restricted by the ethical approvals and the legislation of the European Union.

\clearpage

\bibliographystyle{abbrvnat} 
\bibliography{references.bib}

\begin{thebibliography}{25}
\providecommand{\natexlab}[1]{#1}
\providecommand{\url}[1]{\texttt{#1}}
\expandafter\ifx\csname urlstyle\endcsname\relax
  \providecommand{\doi}[1]{doi: #1}\else
  \providecommand{\doi}{doi: \begingroup \urlstyle{rm}\Url}\fi

\bibitem[Arku et~al.(2020)Arku, Brauer, Ahmed, AlHabib, Avezum, Bo, Choudhury, Dans, Gupta, Iqbal, Ismail, Kelishadi, Khatib, Koon, Kumar, Lanas, Lear, Wei, Lopez-Jaramillo, Mohan, Poirier, Puoane, Rangarajan, Rosengren, Soman, Caklili, Yang, Yeates, Yin, Yusoff, Zatoński, Yusuf, and Hystad]{arku_long-term_2020}
R.~E. Arku, M.~Brauer, S.~H. Ahmed, K.~F. AlHabib, A.~Avezum, J.~Bo, T.~Choudhury, A.~M. Dans, R.~Gupta, R.~Iqbal, N.~Ismail, R.~Kelishadi, R.~Khatib, T.~Koon, R.~Kumar, F.~Lanas, S.~A. Lear, L.~Wei, P.~Lopez-Jaramillo, V.~Mohan, P.~Poirier, T.~Puoane, S.~Rangarajan, A.~Rosengren, B.~Soman, O.~T. Caklili, S.~Yang, K.~Yeates, L.~Yin, K.~Yusoff, T.~Zatoński, S.~Yusuf, and P.~Hystad.
\newblock Long-term exposure to outdoor and household air pollution and blood pressure in the {Prospective} {Urban} and {Rural} {Epidemiological} ({PURE}) study.
\newblock \emph{Environmental Pollution}, 262:\penalty0 114197, July 2020.
\newblock ISSN 02697491.
\newblock \doi{10.1016/j.envpol.2020.114197}.
\newblock URL \url{https://linkinghub.elsevier.com/retrieve/pii/S0269749119335523}.

\bibitem[Chib and Carlin(1999)]{chib_mcmc_1999}
S.~Chib and B.~P. Carlin.
\newblock On {MCMC} sampling in hierarchical longitudinal models.
\newblock \emph{Statistics and Computing}, 9\penalty0 (1):\penalty0 17--26, Apr. 1999.
\newblock ISSN 1573-1375.
\newblock \doi{10.1023/A:1008853808677}.
\newblock URL \url{https://doi.org/10.1023/A:1008853808677}.

\bibitem[Coker et~al.(2018)Coker, Liverani, Su, and Molitor]{coker_multi-pollutant_2018}
E.~Coker, S.~Liverani, J.~G. Su, and J.~Molitor.
\newblock Multi-pollutant {Modeling} {Through} {Examination} of {Susceptible} {Subpopulations} {Using} {Profile} {Regression}.
\newblock \emph{Current Environmental Health Reports}, 5\penalty0 (1):\penalty0 59--69, Mar. 2018.
\newblock ISSN 2196-5412.
\newblock \doi{10.1007/s40572-018-0177-0}.
\newblock URL \url{http://link.springer.com/10.1007/s40572-018-0177-0}.

\bibitem[Dahl and Newton(2007)]{dahl_multiple_2007}
D.~B. Dahl and M.~A. Newton.
\newblock Multiple {Hypothesis} {Testing} by {Clustering} {Treatment} {Effects}.
\newblock \emph{Journal of the American Statistical Association}, 102\penalty0 (478):\penalty0 517--526, June 2007.
\newblock ISSN 0162-1459, 1537-274X.
\newblock \doi{10.1198/016214507000000211}.
\newblock URL \url{http://www.tandfonline.com/doi/abs/10.1198/016214507000000211}.

\bibitem[Fritsch and Ickstadt(2009)]{fritsch_improved_2009}
A.~Fritsch and K.~Ickstadt.
\newblock Improved criteria for clustering based on the posterior similarity matrix.
\newblock \emph{Bayesian Analysis}, 4\penalty0 (2), June 2009.
\newblock ISSN 1936-0975.
\newblock \doi{10.1214/09-BA414}.
\newblock URL \url{https://projecteuclid.org/journals/bayesian-analysis/volume-4/issue-2/Improved-criteria-for-clustering-based-on-the-posterior-similarity-matrix/10.1214/09-BA414.full}.

\bibitem[{GECCO Consortium} et~al.(2020){GECCO Consortium}, Lakerveld, Wagtendonk, Vaartjes, and Karssenberg]{gecco_consortium_deep_2020}
{GECCO Consortium}, J.~Lakerveld, A.~Wagtendonk, I.~Vaartjes, and D.~Karssenberg.
\newblock Deep phenotyping meets big data: the {Geoscience} and {hEalth} {Cohort} {COnsortium} ({GECCO}) data to enable exposome studies in {The} {Netherlands}.
\newblock \emph{International Journal of Health Geographics}, 19\penalty0 (1):\penalty0 49, Dec. 2020.
\newblock ISSN 1476-072X.
\newblock \doi{10.1186/s12942-020-00235-z}.
\newblock URL \url{https://ij-healthgeographics.biomedcentral.com/articles/10.1186/s12942-020-00235-z}.

\bibitem[Griffin et~al.(2022)Griffin, Aristizabal-Henao, Timshina, Ditz, Camacho, Da~Silva, Coker, Deliz~Quiñones, Aufmuth, and Bowden]{griffin_assessment_2022}
E.~K. Griffin, J.~Aristizabal-Henao, A.~Timshina, H.~L. Ditz, C.~G. Camacho, B.~F. Da~Silva, E.~S. Coker, K.~Y. Deliz~Quiñones, J.~Aufmuth, and J.~A. Bowden.
\newblock Assessment of per- and polyfluoroalkyl substances ({PFAS}) in the {Indian} {River} {Lagoon} and {Atlantic} coast of {Brevard} {County}, {FL}, reveals distinct spatial clusters.
\newblock \emph{Chemosphere}, 301:\penalty0 134478, Aug. 2022.
\newblock ISSN 00456535.
\newblock \doi{10.1016/j.chemosphere.2022.134478}.
\newblock URL \url{https://linkinghub.elsevier.com/retrieve/pii/S0045653522009717}.

\bibitem[Hubert and Arabie(1985)]{hubert_comparing_1985}
L.~Hubert and P.~Arabie.
\newblock Comparing partitions.
\newblock \emph{Journal of Classification}, 2\penalty0 (1):\penalty0 193--218, Dec. 1985.
\newblock ISSN 0176-4268, 1432-1343.
\newblock \doi{10.1007/BF01908075}.
\newblock URL \url{http://link.springer.com/10.1007/BF01908075}.

\bibitem[Ishwaran and James(2001)]{ishwaran_gibbs_2001}
H.~Ishwaran and L.~F. James.
\newblock Gibbs {Sampling} {Methods} for {Stick}-{Breaking} {Priors}.
\newblock \emph{Journal of the American Statistical Association}, 96\penalty0 (453):\penalty0 161--173, Mar. 2001.
\newblock ISSN 0162-1459.
\newblock \doi{10.1198/016214501750332758}.
\newblock URL \url{https://doi.org/10.1198/016214501750332758}.
\newblock Publisher: Taylor \& Francis \_eprint: https://doi.org/10.1198/016214501750332758.

\bibitem[Liverani et~al.(2015)Liverani, Hastie, Azizi, Papathomas, and Richardson]{liverani_premium_2015}
S.~Liverani, D.~I. Hastie, L.~Azizi, M.~Papathomas, and S.~Richardson.
\newblock {PReMiuM}: {An} {R} package for profile regression mixture models using {Dirichlet} processes.
\newblock \emph{Journal of statistical software}, 64\penalty0 (7):\penalty0 1, 2015.
\newblock URL \url{https://www.ncbi.nlm.nih.gov/pmc/articles/PMC4905523/}.
\newblock Publisher: Europe PMC Funders.

\bibitem[Medvedovic et~al.(2004)Medvedovic, Yeung, and Bumgarner]{medvedovic_bayesian_2004}
M.~Medvedovic, K.~Yeung, and R.~Bumgarner.
\newblock Bayesian mixture model based clustering of replicated microarray data.
\newblock \emph{Bioinformatics}, 20\penalty0 (8):\penalty0 1222--1232, May 2004.
\newblock ISSN 1367-4811, 1367-4803.
\newblock \doi{10.1093/bioinformatics/bth068}.
\newblock URL \url{https://academic.oup.com/bioinformatics/article/20/8/1222/209795}.

\bibitem[Molitor et~al.(2010)Molitor, Papathomas, Jerrett, and Richardson]{molitor_bayesian_2010}
J.~Molitor, M.~Papathomas, M.~Jerrett, and S.~Richardson.
\newblock Bayesian profile regression with an application to the {National} survey of children's health.
\newblock \emph{Biostatistics}, 11\penalty0 (3):\penalty0 484--498, July 2010.
\newblock ISSN 1468-4357, 1465-4644.
\newblock \doi{10.1093/biostatistics/kxq013}.
\newblock URL \url{https://academic.oup.com/biostatistics/article/11/3/484/257407}.

\bibitem[Murray et~al.(2020)Murray, Aravkin, and {al.}]{murray_global_2020}
C.~J.~L. Murray, A.~Y. Aravkin, and {al.}
\newblock Global burden of 87 risk factors in 204 countries and territories, 1990–2019: a systematic analysis for the {Global} {Burden} of {Disease} {Study} 2019.
\newblock \emph{The Lancet}, 396\penalty0 (10258):\penalty0 1223--1249, 2020.
\newblock ISSN 0140-6736.
\newblock \doi{https://doi.org/10.1016/S0140-6736(20)30752-2}.
\newblock URL \url{https://www.sciencedirect.com/science/article/pii/S0140673620307522}.

\bibitem[Münzel et~al.(2021)Münzel, Sørensen, Lelieveld, Hahad, Al-Kindi, Nieuwenhuijsen, Giles-Corti, Daiber, and Rajagopalan]{munzel_heart_2021}
T.~Münzel, M.~Sørensen, J.~Lelieveld, O.~Hahad, S.~Al-Kindi, M.~Nieuwenhuijsen, B.~Giles-Corti, A.~Daiber, and S.~Rajagopalan.
\newblock Heart healthy cities: genetics loads the gun but the environment pulls the trigger.
\newblock \emph{European Heart Journal}, 42\penalty0 (25):\penalty0 2422--2438, July 2021.
\newblock ISSN 0195-668X.
\newblock \doi{10.1093/eurheartj/ehab235}.
\newblock URL \url{https://doi.org/10.1093/eurheartj/ehab235}.

\bibitem[Neter et~al.(2003)Neter, Stam, Kok, Grobbee, and Geleijnse]{neter_influence_2003}
J.~E. Neter, B.~E. Stam, F.~J. Kok, D.~E. Grobbee, and J.~M. Geleijnse.
\newblock Influence of {Weight} {Reduction} on {Blood} {Pressure}: {A} {Meta}-{Analysis} of {Randomized} {Controlled} {Trials}.
\newblock \emph{Hypertension}, 42\penalty0 (5):\penalty0 878--884, Nov. 2003.
\newblock ISSN 0194-911X, 1524-4563.
\newblock \doi{10.1161/01.HYP.0000094221.86888.AE}.
\newblock URL \url{https://www.ahajournals.org/doi/10.1161/01.HYP.0000094221.86888.AE}.

\bibitem[Nieuwenhuijsen(2018)]{nieuwenhuijsen_influence_2018}
M.~J. Nieuwenhuijsen.
\newblock Influence of urban and transport planning and the city environment on cardiovascular disease.
\newblock \emph{Nature Reviews Cardiology}, 15\penalty0 (7):\penalty0 432--438, July 2018.
\newblock ISSN 1759-5002, 1759-5010.
\newblock \doi{10.1038/s41569-018-0003-2}.
\newblock URL \url{https://www.nature.com/articles/s41569-018-0003-2}.

\bibitem[Oakes et~al.(2014)Oakes, Baxter, and Long]{oakes_evaluating_2014}
M.~Oakes, L.~Baxter, and T.~C. Long.
\newblock Evaluating the application of multipollutant exposure metrics in air pollution health studies.
\newblock \emph{Environment International}, 69:\penalty0 90--99, Aug. 2014.
\newblock ISSN 0160-4120.
\newblock \doi{10.1016/j.envint.2014.03.030}.
\newblock URL \url{https://www.sciencedirect.com/science/article/pii/S016041201400110X}.

\bibitem[Papathomas et~al.(2012)Papathomas, Molitor, Hoggart, Hastie, and Richardson]{papathomas_exploring_2012}
M.~Papathomas, J.~Molitor, C.~Hoggart, D.~Hastie, and S.~Richardson.
\newblock Exploring data from genetic association studies using {Bayesian} variable selection and the {Dirichlet} process: application to searching for gene × gene patterns.
\newblock \emph{Genetic Epidemiology}, 36\penalty0 (6):\penalty0 663--674, Sept. 2012.
\newblock ISSN 1098-2272.
\newblock \doi{10.1002/gepi.21661}.

\bibitem[Pitman(2006)]{pitman_combinatorial_2006}
J.~Pitman.
\newblock \emph{Combinatorial stochastic processes: {Ecole} d'eté de probabilités de saint-flour xxxii-2002}.
\newblock Springer, 2006.
\newblock URL \url{https://books.google.com/books?hl=nl&lr=&id=mW32BwAAQBAJ&oi=fnd&pg=PA1&dq=pitman+Combinatorial+stochastic+processes.&ots=j17lC74GmN&sig=Juf0S3qGtxSC_48pIPIyFxO6OSo}.

\bibitem[Reckelhoff(2001)]{reckelhoff_gender_2001}
J.~F. Reckelhoff.
\newblock Gender {Differences} in the {Regulation} of {Blood} {Pressure}.
\newblock \emph{Hypertension}, 37\penalty0 (5):\penalty0 1199--1208, May 2001.
\newblock ISSN 0194-911X, 1524-4563.
\newblock \doi{10.1161/01.HYP.37.5.1199}.
\newblock URL \url{https://www.ahajournals.org/doi/10.1161/01.HYP.37.5.1199}.

\bibitem[Rouanet et~al.(2023)Rouanet, Johnson, Strauss, Richardson, Tom, White, and Kirk]{rouanet_bayesian_2023}
A.~Rouanet, R.~Johnson, M.~Strauss, S.~Richardson, B.~D. Tom, S.~R. White, and P.~D.~W. Kirk.
\newblock Bayesian profile regression for clustering analysis involving a longitudinal response and explanatory variables.
\newblock \emph{Journal of the Royal Statistical Society Series C: Applied Statistics}, page qlad097, Nov. 2023.
\newblock ISSN 0035-9254, 1467-9876.
\newblock \doi{10.1093/jrsssc/qlad097}.
\newblock URL \url{https://academic.oup.com/jrsssc/advance-article/doi/10.1093/jrsssc/qlad097/7382190}.

\bibitem[Scrucca et~al.(2016)Scrucca, Fop, Murphy, and Raftery]{scrucca_mclust_2016}
L.~Scrucca, M.~Fop, T.~B. Murphy, and A.~E. Raftery.
\newblock mclust 5: {Clustering}, {Classification} and {Density} {Estimation} {Using} {Gaussian} {Finite} {Mixture} {Models}.
\newblock \emph{The R Journal}, 8\penalty0 (1):\penalty0 289--317, Aug. 2016.
\newblock ISSN 2073-4859.

\bibitem[Sijtsma et~al.(2022)Sijtsma, Rienks, Van Der~Harst, Navis, Rosmalen, and Dotinga]{sijtsma_cohort_2022}
A.~Sijtsma, J.~Rienks, P.~Van Der~Harst, G.~Navis, J.~G.~M. Rosmalen, and A.~Dotinga.
\newblock Cohort {Profile} {Update}: {Lifelines}, a three-generation cohort study and biobank.
\newblock \emph{International Journal of Epidemiology}, 51\penalty0 (5):\penalty0 e295--e302, Oct. 2022.
\newblock ISSN 0300-5771, 1464-3685.
\newblock \doi{10.1093/ije/dyab257}.
\newblock URL \url{https://academic.oup.com/ije/article/51/5/e295/6460206}.

\bibitem[Vishram et~al.(2012)Vishram, Borglykke, Andreasen, Jeppesen, Ibsen, Jørgensen, Broda, Palmieri, Giampaoli, Donfrancesco, Kee, Mancia, Cesana, Kuulasmaa, Sans, and Olsen]{vishram_impact_2012}
J.~K. Vishram, A.~Borglykke, A.~H. Andreasen, J.~Jeppesen, H.~Ibsen, T.~Jørgensen, G.~Broda, L.~Palmieri, S.~Giampaoli, C.~Donfrancesco, F.~Kee, G.~Mancia, G.~Cesana, K.~Kuulasmaa, S.~Sans, and M.~H. Olsen.
\newblock Impact of {Age} on the {Importance} of {Systolic} and {Diastolic} {Blood} {Pressures} for {Stroke} {Risk}: {The} {MOnica}, {Risk}, {Genetics}, {Archiving}, and {Monograph} ({MORGAM}) {Project}.
\newblock \emph{Hypertension}, 60\penalty0 (5):\penalty0 1117--1123, Nov. 2012.
\newblock ISSN 0194-911X, 1524-4563.
\newblock \doi{10.1161/HYPERTENSIONAHA.112.201400}.
\newblock URL \url{https://www.ahajournals.org/doi/10.1161/HYPERTENSIONAHA.112.201400}.

\bibitem[Wild(2005)]{wild_complementing_2005}
C.~P. Wild.
\newblock Complementing the {Genome} with an “{Exposome}”: {The} {Outstanding} {Challenge} of {Environmental} {Exposure} {Measurement} in {Molecular} {Epidemiology}.
\newblock \emph{Cancer Epidemiology, Biomarkers \& Prevention}, 14\penalty0 (8):\penalty0 1847--1850, Aug. 2005.
\newblock ISSN 1055-9965, 1538-7755.
\newblock \doi{10.1158/1055-9965.EPI-05-0456}.
\newblock URL \url{https://aacrjournals.org/cebp/article/14/8/1847/258124/Complementing-the-Genome-with-an-Exposome-The}.

\end{thebibliography}
\clearpage
\section{Appendix}
\subsection{Gibbs sampler, derivation details.}
In this section, we provide a detailed description of our Gibbs sampling algorithm. Our objective is to sample the joint posterior distribution of the following set of random variables: $\bm{\Theta} = \{\bm{\beta},\sigma,\bm{\gamma},\boldsymbol{W}^\text{Int},\bm{\eta},\boldsymbol{W}^\text{RE},\bm{\theta}^\text{U},\boldsymbol{Z},\bm{\pi},\zeta\}$ where $\bm{\gamma}$ and $\bm{\eta}$ contain the $C$ latent cluster effect vectors and the $m$ individual-specific random effects vectors, respectively.
The pseudo-code below outlines the overall structure of our Gibbs sampler.\\

\begin{algorithm}[!h]
\caption{Profile-LMM Gibbs sampler}\label{alg:GS}
\begin{algorithmic}
\Require $K \geq 0$, $\Tilde{\bm{\Theta}}$ \Comment{Number of samples and initial parameter}
\State $\bm{\Theta}_0 \gets \Tilde{\bm{\Theta}}$ 
\For{$k\gets1$ to $K$}\\
\vspace{-0.15cm}
\State \textit{a) Sample the cluster structure parameters;}
\vspace{0.15cm}
    \State \hspace{0.6cm}$\bm{\theta}^\text{U}_{k} \sim p(\bm{\theta}^\text{U}|\boldsymbol{U},\boldsymbol{Z}_{k-1})$
\vspace{0.35cm}
\State \textit{b) Sample the LMM parameters;}
\vspace{0.15cm}
    \State \hspace{0.6cm} $(\bm{\beta}_k,\bm{\gamma}_{k}) \sim p(\bm{\beta},\bm{\gamma}|\mathbf{y},\boldsymbol{Z}_{k-1},\sigma_{k-1},\bm{\eta}_{k-1},\boldsymbol{W}^\text{Int}_{k-1})$
\vspace{0.05cm}
    \State \hspace{0.6cm} $\sigma_{k} \sim p(\sigma|\mathbf{y},\boldsymbol{Z}_{k-1},\bm{\beta}_k,\bm{\gamma}_{k},\bm{\eta}_{k-1})$
\vspace{0.05cm}
    \State \hspace{0.6cm} $\bm{\eta}_k \sim p(\bm{\eta}|\mathbf{y},\boldsymbol{Z}_{k-1},\bm{\beta}_k,\sigma_{k},\bm{\gamma}_{k},\boldsymbol{W}^\text{RE}_{k-1})$
\vspace{0.05cm}
    \State \hspace{0.6cm} $\boldsymbol{W}^\text{RE}_k \sim p(\boldsymbol{W}^\text{RE}|\bm{\eta}_{k})$
\vspace{0.05cm}
    \State \hspace{0.6cm} $\boldsymbol{W}^\text{Int}_k \sim p(\boldsymbol{W}^\text{Int}|\bm{\gamma}_{k})$
\vspace{0.35cm}
\State \textit{c) Sample the assignment parameters;}
\vspace{0.15cm}
    \State \hspace{0.6cm} $\boldsymbol{Z}_k \sim p(\boldsymbol{Z}|\mathbf{y},\boldsymbol{U},\bm{\beta}_k,\sigma_{k},\bm{\gamma}_{k},\bm{\eta}_k,\bm{\theta}^\text{U}_{k},\bm{\pi}_{k-1})$
\vspace{0.05cm}
    \State \hspace{0.6cm} $\bm{\pi}_k \sim p(\bm{\pi}|\boldsymbol{Z}_k)$
\vspace{0.05cm}
    \State \hspace{0.6cm} $\zeta_k \sim p(\zeta|\bm{\pi}_k)$
\vspace{0.25cm}
\EndFor
\end{algorithmic}
\end{algorithm}
We now explicit the conditional distributions of each block of the Gibbs sampler, and drop the index $k$ for sake of readability.\\

\textbf{\textit{a) Cluster structure parameters conditional distribution}}\\
The prior distribution for the clustering parameters of cluster $c$, $\bm{\theta}^\text{U}_c = (\bm{\mu}_c, \sigma_c)$, is specified as a Normal-Inverse-Wishart distribution, $\text{NIW}(0,\lambda_0,\nu_0,\boldsymbol{\Phi}_0)$.

\noindent The posterior distribution is as follows:
$$
\bm{\theta}^\text{U}_{c}|\boldsymbol{U},\boldsymbol{Z}\sim\text{NIW}\left(\frac{n^c\Bar{\mathbf{U}}_c}{\lambda_0+n^c},\lambda_0+n^c,\nu_0+n^c,\boldsymbol{\Phi}_n\right),
$$
where $\boldsymbol{U}_c$ denotes the rows of $\boldsymbol{U}$ assigned to cluster $c$ (i.e., $\boldsymbol{Z}=c$), $n_c$ is the number of observations in cluster $c$, and $\Bar{\mathbf{U}}_c = \frac{1}{n^c}\sum \mathbf{U}_{c,j}$ is the sample mean of the observations within that cluster. The updated scale matrix is given by:
$$
\boldsymbol{\Phi}_n = \boldsymbol{\Phi}_0 + \mathbf{U}^{\top}_c\mathbf{U}_c-n_c\Bar{\mathbf{U}}_{c}^\top\Bar{\mathbf{U}}_c.
$$
In the case where no observations are allocated to cluster $c$, the posterior distribution reverts to the prior.\\

\textbf{\textit{b)LMM parameters conditional distribution}}\\
For the majority of the LMM parameters, we can sequentially update their posterior distributions using standard conditional distributions. However, as noted by \cite{chib_mcmc_1999}, the fixed effect parameters ($\bm{\beta}$) and the cluster effects ($\bm{\gamma}$) present a common Gibbs sampling challenge in hierarchical models: they exhibit high a-posteriori correlation when they are applied to the same covariates.

To account for this correlation, and following Algorithm 2 from \cite{chib_mcmc_1999}, we sample from their joint posterior distribution. This is done sequentially by marginalising out $\bm{\gamma}$:
$p(\bm{\beta},\bm{\gamma}|\mathbf{y},\boldsymbol{Z},\sigma,\bm{\eta},\boldsymbol{W}^\text{Int}) = p(\bm{\beta}|\mathbf{y},\boldsymbol{Z},\sigma,\bm{\eta},\boldsymbol{W}^\text{Int}) p(\bm{\gamma}|\mathbf{y},\boldsymbol{Z},\bm{\beta},\sigma,\bm{\eta},\boldsymbol{W}^\text{Int})$.
This approach is tractable because the conditional distribution of $\bm{\beta}$ while integrating out $\bm{\gamma}$ has a closed form, given the conjugate prior $\bm{\beta}\sim\mathcal{N}(\bm{0},\frac{\sigma^2}{\lambda}\mathbf{I})$:
$$
\bm{\beta}|\mathbf{y},\boldsymbol{Z},\sigma,\bm{\eta},\boldsymbol{W}^\text{Int} \sim \mathcal{N}(\hat{\boldsymbol{\beta}},\mathbf{B}),
$$
where $\hat{\boldsymbol{\beta}} = \mathbf{B}\sum_{c=1}^C\mathbf{x}^{\text{Fe}\top}_c\mathbf{V}_c^{-1}(\mathbf{y}_c-\mathbf{y}^{\text{Re}}_c)$ and $\mathbf{B} =\left( \frac{\lambda}{\sigma^2}\mathbf{I}+\sum_{c=1}^C\mathbf{x}^{\text{Fe}\top}_c\mathbf{V}^{-1}_c\mathbf{x}^{\text{Fe}}_c\right)^{-1}$. The term $\mathbf{V}_c = \sigma^2 \mathbf{I}+\mathbf{x}^{\text{Int}\top}_c\boldsymbol{W}^\text{Int}\mathbf{x}^{\text{Int}}_c$ and $\mathbf{y}^{\text{Re}} = \mathbf{x}^{\text{Re}}\bm{\eta}$ is the random effects contribution to the outcome. The subscript $c$ indicates that the matrices are restricted to the rows corresponding to observations where $\boldsymbol{Z}=c$.\\
We can then turn our attention to the conditional posterior distribution of $\bm{\gamma}$:
$$
\bm{\gamma}|\boldsymbol{Y},\boldsymbol{Z},\bm{\beta},\sigma,\bm{\eta},\boldsymbol{W}^\text{Int}\sim \mathcal{N}(\hat{\boldsymbol{\mu}}_c,\mathbf{D}_c),
$$
where $\hat{\boldsymbol{\mu}}_c = \mathbf{x}^{\text{Int}\top}_c\mathbf{V}_c^{-1}\mathbf{x}^{\text{Int}}_c(\mathbf{y}_c-\mathbf{y}^{\text{Re}}_c-\mathbf{x}^{\text{Fe}}_c\boldsymbol{\beta})$ and $\mathbf{D}_c =\boldsymbol{W}^\text{Int} - \boldsymbol{W}^\text{Int}\mathbf{x}^{\text{Int}\top}_c\mathbf{V}_c^{-1}\mathbf{x}^{\text{Int}}_c\boldsymbol{W}^\text{Int} $.\\
The remainder of the conditionals are more straightforward given the conjugated nature of the priors. The prior of the error variance $\sigma^2$ is an Gamma distribution $\sigma^2\sim \mathcal{G}(a,b)$ and it's conditional posterior distribution is :
$$
\sigma|\mathbf{y},\boldsymbol{Z},\bm{\beta},\bm{\gamma},\bm{\eta}\sim\mathcal{G} \left (a+\frac{n}{2},b+\frac{1}{2}\sum_{i=1}^n(\mathbf{x}^{\mathbf{y}_i -\text{Fe}}_i\bm{\beta} - \mathbf{x}^{\text{Re}}_i\bm{\eta}_{g(i)}-\mathbf{x}^{\text{Int}}_i\bm{\gamma}_{Z_i})^2\right).
$$
The random effect coefficient for individual $j$ conditional posterior's distribution is :
$$
\bm{\eta}_j|\mathbf{y},\boldsymbol{Z},\bm{\beta},\sigma,\bm{\gamma},\boldsymbol{W}^\text{RE}\sim \mathcal{N}(\hat{\boldsymbol{\eta}}_j,\mathbf{E}_j),
$$
where $\hat{\boldsymbol{\eta}}_j = \mathbf{x}^{\text{Re}\top}_j\mathbf{S}_j^{-1}\mathbf{x}^{\text{Re}}_j(\mathbf{y}_j-\mathbf{y}^{\text{Int}}_j-\mathbf{x}^{\text{Fe}}_j\boldsymbol{\beta})$ and $\mathbf{E}_j =\boldsymbol{W}^\text{Re} - \boldsymbol{W}^\text{Re}\mathbf{x}^{\text{Re}\top}_j\mathbf{S}_j^{-1}\mathbf{x}^{\text{Re}}_j\boldsymbol{W}^\text{Re} $. The $\mathbf{y}^{\text{Int}} = \mathbf{x}^{\text{Int}}\bm{\gamma}$ is the contribution to the outcome of the profile cluster component and $\mathbf{S}_j = \sigma^2 \mathbf{I}+\mathbf{x}^{\text{Re}\top}_j\boldsymbol{W}^\text{Re}\mathbf{x}^{\text{Re}}_j$. Similarly to the interaction term the subscript $j$ indicates that the matrices are restricted to the rows corresponding to individual $j$ (where $g(i)=j$). \\
The prior of the individual random effects covariance matrix $\boldsymbol{W}^\text{RE}$, is an inverse-Wishart distribution $\boldsymbol{W}^\text{RE}\sim\mathcal{IW}(\bm{\Psi}^{\text{RE}},\nu^{\text{RE}})$ and its conditional posterior's distribution is :
$$
\boldsymbol{W}^\text{RE}|\bm{\eta}\sim\mathcal{IW}(\bm{\Psi}^{\text{RE}}+\bm{\eta}\bm{\eta}^\top,\nu^{\text{RE}}+m).
$$
Similarly for the variance of the cluster interaction coefficients:
$$\boldsymbol{W}^\text{Int}|\bm{\gamma}\sim\mathcal{IW}(\bm{\Psi}^{\text{Int}}+\bm{\gamma}\bm{\gamma}^\top,\nu^{\text{Int}}+C),
$$
where $C$ is the maximal number of clusters considered.\\

\textbf{\textit{c) Cluster structure parameters conditional distribution}}\\
For each observation $i$, we draw its cluster belonging $\boldsymbol{Z}_{k,i}$ from a $C$ dimensional multinomial distribution $\boldsymbol{Z}_{k,i}\sim\mathcal{M}(p_1,\cdots,p_C)$, where each weight $p_c$ is given by :
$$
p_c \propto \bm{\pi}_c p( \mathbf{y}_i|\bm{\beta},\sigma,\bm{\gamma}_c,\bm{\eta})
p( \boldsymbol{U}_i|\bm{\theta}^\text{U}_{c}).
$$
Following \cite{ishwaran_gibbs_2001}, we use the following procedure to draw the weights $\bm{\pi}$ from their marginal. First, draw $C$ auxilary variables $V_c$ from a beta distribution :
\begin{align*}
    V_c\sim\beta(1+n_c,\zeta+\sum_{l=c+1}^Cn_l),
\end{align*}
where $n_c$ is the number of observations that are assigned to cluster $c$. The weights are then defined as:
$$
\pi_1 = V_1,\,\,\,\,\pi_c = V_c\prod_{l=1}^{c-1}(1-V_l).
$$
Finally, the parameter $\zeta$, to which a Gamma prior distribution is assigned, $\zeta\sim\text{Gamma}(a,b)$ has the following marginal posterior distribution:
$$
\zeta|\bm{\pi}\sim \text{Gamma}(a+C,o),
$$
where $o  = \left(b^{-1} - \sum_{i=1}^n \ln(1-V_i)\right)^{-1}$. The $o$ is defined in terms of the auxiliary variables $V_i$, which can be retrieved from the values of $\bm{\pi}$.
\clearpage
\section{Supplementary material}
\subsection{Simulation results}


\clearpage

\subsection{Real data exposure correlation and MCMC chain convergence.}
\begin{table}[h!]
\centering
\begin{tabular}{|c|cccc|}
  \hline
            &  NO2&       OZO  &     P25 &  NDV\_MD3 \\ 
  \hline
NO2    & 1.00 &0.71& 0.82& 0.19 \\ 
OZO     &0.71 &1.00& 0.91& 0.69\\
P25    & 0.82& 0.91& 1.00& 0.54\\
NDV\_MD3 &0.19& 0.69& 0.54& 1.00\\
   \hline
\end{tabular}
\caption{Correlation matrix of the exposures.} 
\label{tab:corExp}
\end{table}

\begin{figure} [h!]
    \centering
    \includegraphics[width=1\linewidth]{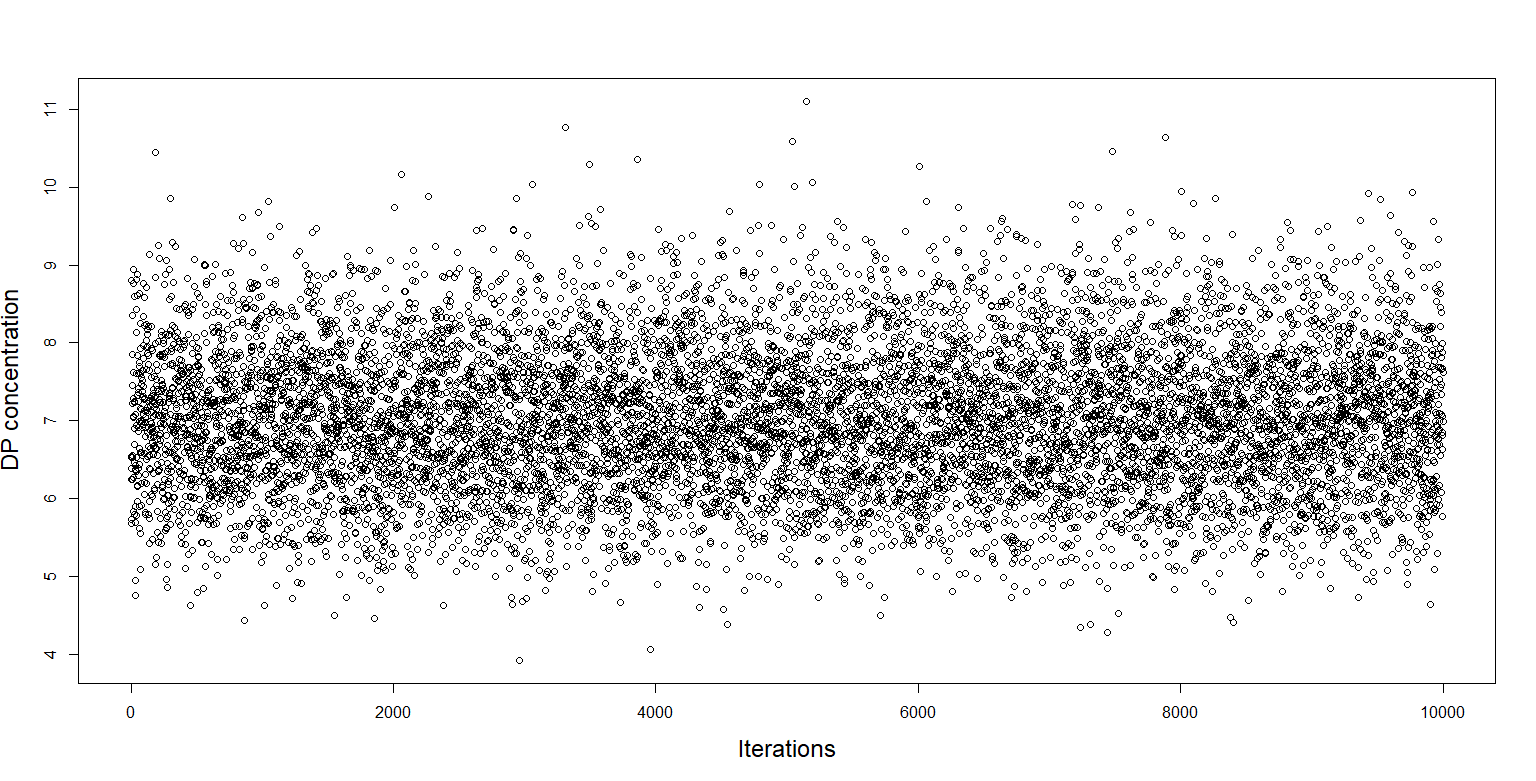}
    \caption{Trace of the DP concentration parameter after a burn-in of 5,000 iterations. }
    \label{fig:conc}
\end{figure}

\clearpage
\subsection{Real data validation dataset results.}

\begin{table}[h!]
\centering
\begin{tabular}{|c|ccccc|}
  \hline
Cluster  & 1 & 2 & 3 & 4 & 5  \\ 
  \hline
Nb. observations & 2334 & 464 & 1073 & 923 & 5206 \\ 
   \hline
\end{tabular}
\caption{Distribution of the 10,000 observations among the 5 clusters identified by the profile LMM. Estimation done on a different subset of the data.} 
\label{tab:ClusStab}
\end{table}

\begin{figure}[h!]
    \centering
    \includegraphics[width=1\linewidth]{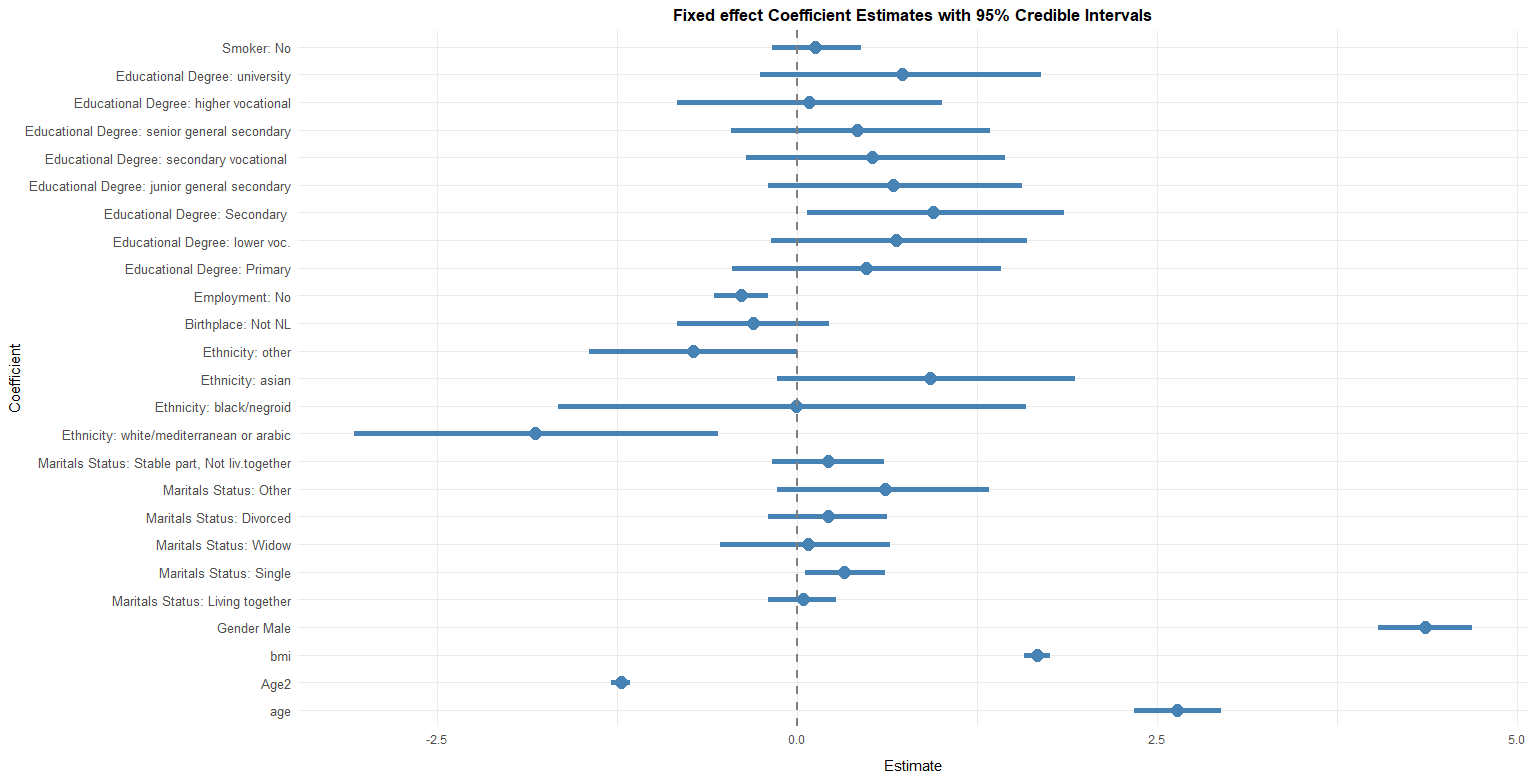}
    \caption{Fixed effect estimates with 95\% credible intervals of standardized coefficients, DBP as an outcome. Estimation done on a different subset of the data.}
    \label{fig:FERealStab}
\end{figure}

\begin{figure}[h!]
    \centering
    \includegraphics[width=1\linewidth]{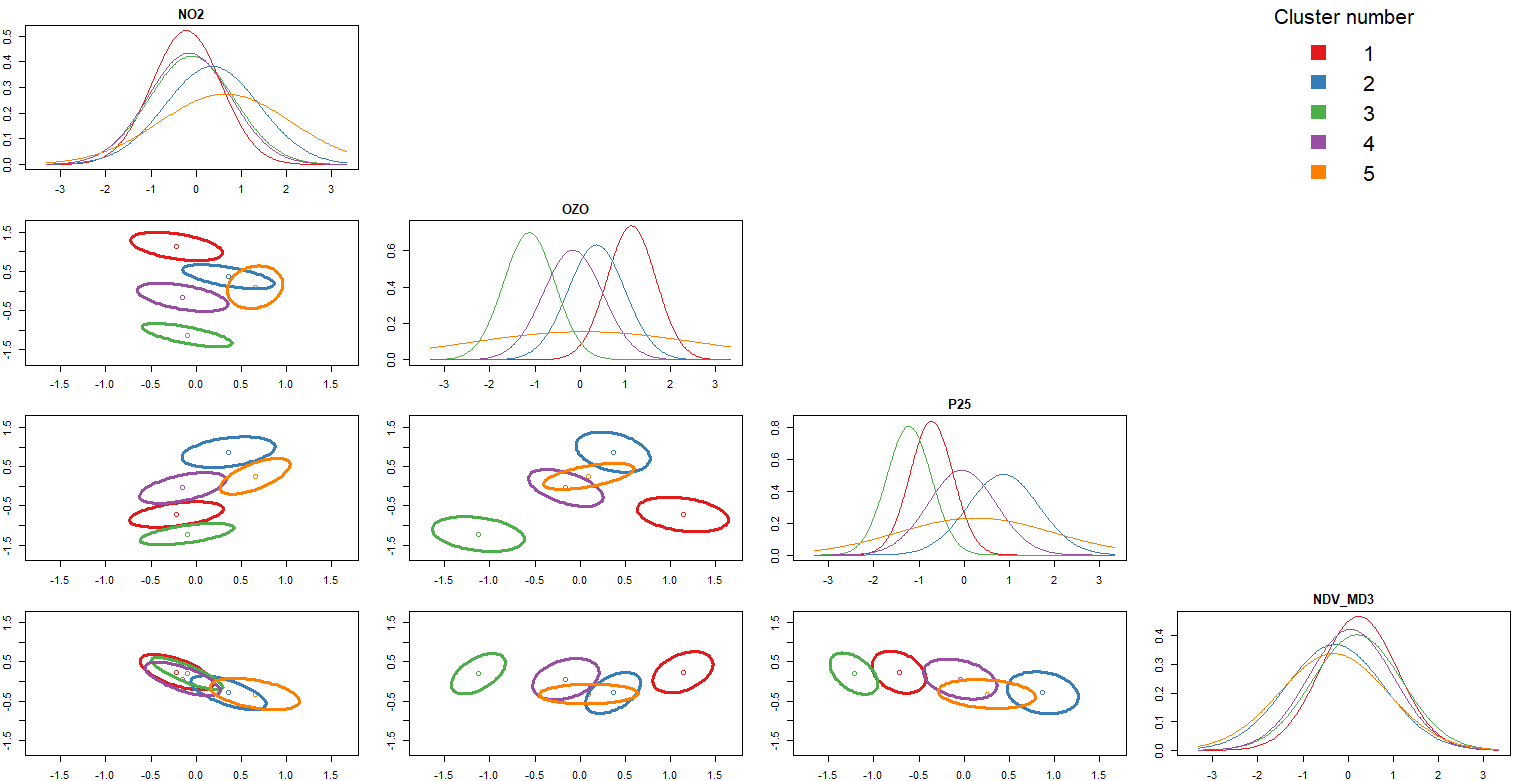}
    \caption{Exposure latent clusters mixtures representations. Diagonal panels show the marginals for each exposure and off-diagonal panels the pairwise marginal distributions. The dot indicates the centroid of the mixture and the ellipse is a representation of the covariance matrix. Estimation done on a different subset of the data. }
    \label{fig:clusterRealStab}
\end{figure}

\clearpage

\end{document}